\documentclass[
 reprint,
 superscriptaddress,
 amsmath,amssymb,
 jcp,
 floatfix,
]{revtex4-2}
\usepackage{graphicx}
\usepackage{dcolumn}
\usepackage{bm}
\usepackage[usenames,dvipsnames]{xcolor} 
\usepackage[
	open,
	openlevel=2,
	atend
]{bookmark}
\usepackage[]{hyperref}            
\usepackage{tikz}
\usetikzlibrary{arrows}
\usepackage{flowchart}

\newcommand{\rstate}[1]{\vert #1 \rangle}

\newcommand{\avstate}[3]{\langle #1 \vert #2 \vert #3 \rangle }
\newcommand{\dotstate}[2]{\langle #1 \vert #2  \rangle }

\newcommand{\tf}[1]{\text{#1}}
\newcommand{\tb}[1]{\textbf{#1}}
\newcommand{\ti}[1]{\textit{#1}}
\newcommand{\etal}{\textit{et al.} }
\newcommand{\etalcite}{\textit{et al.}}

\begin{document}
\title{Self-interaction corrected SCAN functional for molecules and solids in the numeric atom-center orbital framework}
\newcommand{\addriChEM}{Collaborative Innovation Center of Chemistry for Energy Materials}
\newcommand{\addrSHMCIM}{Shanghai Key Laboratory of Molecular Catalysis and Innovative Materials}
\newcommand{\addrMOECPS}{MOE Key Laboratory of Computational Physical Sciences, Shanghai Key Laboratory of Bioactive Small Molecules}
\newcommand{\addrSKLBSM}{Shanghai Key Laboratory of Bioactive Small Molecules}
\newcommand{\addrFudan}{Department of Chemistry, Fudan University, Shanghai 200433, People's Republic of China}
\newcommand{\addrHFNL}{Hefei National Laboratory, Hefei 230088, China}
\newcommand{\addrZJLab}{Research Center for Intelligent Supercomputing, Zhejiang Lab, Hangzhou 311100, P. R. China}
\newcommand{\addrFHI}{The NOMAD Laboratory at the FHI of the Max-Planck-Gesellschaft and IRIS-Adlershof of the Humboldt-Universit\"{a}t zu Berlin, Faradayweg 4-6, D-14195 Berlin-Dahlem Germany}
\author{Sheng Bi}
\affiliation{\addrFHI}
\affiliation{\addrFudan}
\affiliation{\addrZJLab}
\author{Christian Carbogno}
\email{christian.carbogno@fhi-berlin.mpg.de}
\affiliation{\addrFHI}	
\author{Igor Ying Zhang}
\email{igor\_zhangying@fudan.edu.cn}
\affiliation{\addrFudan}
\affiliation{\addrMOECPS}
\author{Matthias Scheffler}
\affiliation{\addrFHI}	
\date{\today}
\begin{abstract}
	Semilocal density-functional approximations (DFAs), including the state-of-the-art SCAN functional, are plagued by the self-interaction error (SIE). While this error is explicitly defined only for one-electron systems, it has inspired the self-interaction correction method proposed by Perdew and Zunger (PZ-SIC), which has shown promise in mitigating the many-electron SIE. However, the PZ-SIC method is known for its significant numerical instability. 
	In this study, we introduce a novel constraint that facilitates self-consistent localization of the SIC orbitals in the spirit of Edmiston-Ruedenberg orbitals [Rev. Mod. Phys. \textbf{35}, 457 (1963)]. 
	Our practical implementation within the all-electron numeric atom-centered orbitals code \textit{FHI-aims} guarantees efficient and stable convergence of the self-consistent PZ-SIC equations for both molecules and solids. We further demonstrate that our PZ-SIC approach effectively mitigates the SIE in the meta-GGA SCAN functional, significantly improving the accuracy for ionization potentials, charge-transfer energies, and band gaps for a diverse selection of molecules and solids. However, our PZ-SIC method does have its limitations. It can not improve the already accurate SCAN results for properties such as cohesive energies, lattice constants, and bulk modulus in our test sets. This highlights the need for new-generation DFAs with more comprehensive applicability.  

\end{abstract}
                              
\maketitle

\section{Introduction}
Kohn-Sham (KS) density-functional theory (DFT)~\cite{cite-GEA-1,cite-GEA-2} is currently the most widely used electronic-structure method across various scientific disciplines \cite{comput-2022-data-driven, DFT-numerical-test,DFT-latest-2022}. 
To a large extent, this is due to the favorable balance between computational accuracy and efficiency. The key contribution to this favorable balance is the existence of density-functional approximations (DFAs) to the exact exchange-correlation (XC) functional~\cite{cite-GEA-2, DFT-latest-2022}. In this context, the non-empirical SCAN functional proposed by Sun, Ruzsinszky, and Perdew in 2015~\cite{cite-scan-1} is particularly promising. Compared to earlier semi-local DFAs, including the local-density approximations (LDAs), the generalized gradient approximations (GGAs), and other meta-GGAs~\cite{cite-meta-1,cite-meta-2,cite-meta-3}, the SCAN meta-GGA functional yields consistent and notable improvements in describing many chemical and physical properties of both molecules and solids~\cite{cite-scan-2-ferroelectric,cite-scan-3-vdW,cite-scan-molecule}. This can be attributed to its compliance with 17 constraints, which the exact XC functional must satisfy.
However, like all other semi-local DFAs, the SCAN functional is not immune to the self-interaction error (SIE)~\cite{cite-DFT-SIE, cite-SIC-intro,cite-SIC-4-manybody}. This limitation accounts for its inability to reliably describe the ionization potential, specifically the energy of the highest occupied molecular orbital, as well as charge-transfer properties and band gaps in insulators and semiconductors~\cite{cite-DFT-challenge,cite-scan-3-solid}.

Formally, the SIE arises from the ground of the XC functional within the KS-DFT framework, i.e., the Hartree term \footnote{Hartree atomic units are used throughout the paper.}: 
\begin{equation} 
  \begin{aligned}\label{Hartree}
    E_{\text{H}}[n] := \frac{1}{2} \iint \frac{n(\textbf{r}) n(\textbf{r}')}{| \textbf{r} - \textbf{r}'|} d^3  \textbf{r}' d^3  \textbf{r}\quad,
  \end{aligned}
\end{equation}
which captures the most part of electron-electron Coulomb interaction in the system. 
However, since the density $n(\textbf{r})$ encompasses all electrons, the Hartee term also introduces non-physical Coulomb interactions where individual electrons interact with themselves. 
In one-electron systems, this spurious interaction is precisely counterbalanced by the Fock term within the Hartree-Fock (HF) theory. In other words, the HF method is free from one-electron self-interactions. However, it is well-documented that the HF approach is not enough to cancel out these non-physical self-interactions in multi-electron systems \cite{cite-delocalized-error}. Within the KS-DFT framework, these non-physical interactions are expected to be precisely canceled out by the exact XC functional. 

Unfortunately, this cancellation is insufficiently achieved for all existing semi-local DFAs. The residual self-interaction implies the SIE of a given DFA, leading to incorrect electron delocalization~\cite{cite-DFT-SIE, cite-SIC-gap}. 
Obviously, a single orbital's XC energy must cancel its self-interacted Coulomb interaction. 
This analysis has motivated the development of so-called self-interaction corrections~(SIC), initially proposed by Perdew and Zunger in the early 1980s (PZ-SIC)~\cite{cite-SIC-4-manybody}:
\begin{equation}
  \begin{aligned} \label{eq:SIC}
	  E^{\text{PZ-SIC}}[\{n_i\}] := - \sum_i^{N_{\text{e}}} \left( E_{\text{xc}}^{\text{DFA}}[n_{i}]+E_{\text{H}}[n_{i}] \right) \quad.
  \end{aligned}
\end{equation}
Here, $N_e$ represents the total number of electrons in the system, and $\{ n_i(\textbf{r}) = |\phi_i (\textbf{r})|^2 \}$ denotes the densities of occupied single-electron orbitals, referred to as SIC orbitals~$ \{\phi_i (\textbf{r})\}$ henceforth (for further details, see Section \ref{sec-theory}). Accordingly, $E_{\text{xc}}^{\text{DFA}}[n_{i}]$ is the DFA-specific single-particle XC energy evaluated for the SIC orbital density $n_i(\textbf{r})$. Meanwhile, $E_{\text{H}}[n_{i}]$ is the Coulomb self-interaction energy of $n_i(\textbf{r})$, calculated using the Hartree equation~(Eq.~(\ref{Hartree})). By incorporating these one-electron correction terms into the semi-local DFA XC energy, the PZ-SIC scheme effectively eliminates the SIE of a given DFA in one-electron systems. Meanwhile, it has been shown that the PZ-SIC scheme can substantially mitigate many-electron SIE when the appropriate SIC orbitals are employed~\cite{cite-SIC-25,cite-SIC-25-wannier,cite-SIC-max-wannier}. 

While SIC orbitals coincide with KS orbitals in one-electron systems, they generally diverge in many-electron systems. Theoretically, the optimal SIC orbitals should be determined by variationally minimizing the total energy of the self-interaction corrected DFA~(SIC-DFA)~\cite{cite-SIC-4-manybody}. This minimization must adhere to the \textit{total-density constraint}, which stipulates that the sum of all occupied SIC orbital densities should yield the total electron density~$n(\textbf{r})=\sum_i^{N_{\text{e}}}n_i(\textbf{r})$ (for further details, see Section \ref{sec-theory}). Moreover, as initially proposed by Pederson, Heaton, and Lin, the variational minimization of the SIC-DFA total energy not only produces the PZ-SIC equations for KS orbitals but also imposes the so-called \textit{orbital potential constraint} on SIC orbitals~\cite{cite-SIC-16}. Similar to the standard KS equations, PZ-SIC equations should be solved iteratively in a self-consistent manner because they depend on both KS and SIC orbitals (also see Section \ref{sec-theory} for more details). 

In practice, the PZ-SIC equations often yield multiple self-consistent solutions for SIC orbitals, all satisfying the total density and orbital potential constraints. As a result, while the correct SIC orbitals that minimize the SIC-DFA total energy are among these multiple self-consistent solutions, the specific solution obtained is highly sensitive to the choice of initial guess. 

In the seminal work of PZ-SIC, Perdew and Zunger observed that localized orbitals often provide lower SIC-DFA total energies compared to their delocalized (canonical) counterparts, such as standard KS orbitals~\cite{cite-SIC-4-manybody}. This observation led to frequently characterizing SIC orbitals as ``localized orbitals''.
This insight has inspired several empirical guidelines for initializing SIC orbitals~\cite{cite-SIC-8-lsda}. A variety of localization methods have been used to generate localized initializations~\cite{cite-SIC-complex-multiple}, including Pipek-Mezey~\cite{cite-localized-method-1}, Edmiston-Ruedenberg~\cite{cite-localized-method-2}, von Niessen~\cite{cite-localized-method-3}, Foster-Boys~\cite{cite-localized-method-4}, and the fourth-moment~\cite{cite-localized-method-5} methods. Lehtola, Head-Gordon, and J\'onsson~\cite{cite-SIC-complex-multiple} systematically examined various initial SIC orbitals for atoms in the first three rows of the periodic table. Remarkably, the SIC total energies associated with these different sets of orbitals can diverge by more than 0.5 eV. Consequently, identifying the correct SIC orbitals remains a significant challenge, even for atoms and simple diatomic molecules~\cite{cite-SIC-8-lsda,cite-SIC-16}.

More recently, Pederson, Ruzsinszky, and Perdew noted the topological similarity between Fermi orbitals with L{\"o}wdin orthogonalization (FLOs) and previously employed localized orbitals. Therefore, they suggested constructing the SIC orbitals using FLOs~\cite{cite-SIC-18-Fermi-Theory,cite-SIC-Fermi-detail,cite-SIC-Fermi-symmetry,cite-SIC-20-FullFermi}, giving rise to a modified PZ-SIC algorithm known as FLO-SIC. This innovative approach effectively filters out the FLO-like SIC orbitals from the multiple self-consistent solutions, mitigating the issue of multiple solutions. FLO-SIC has demonstrated considerable promise for atoms and molecules~\cite{cite-SIC-3-fermi-orbital, FLO-Shrinking, FLO-Scaling}, with particular success observed in systems containing $\pi$ bonds~\cite{cite-SIC-3-fermi-orbital,cite-SIC-21-Fermi-C60}.

For extended materials, finding the correct SIC orbitals becomes even more serious~\cite{cite-SIC-off-diagonal}. To capture the localized nature of SIC orbitals, Heaton, Harrison, and Lin suggested representing them in terms of real-space Wannier orbitals~\cite{cite-SIC-25-wannier} under periodic boundary conditions (PBCs). However, it does not thoroughly solve the issue of multiple solutions. Different sets of Wannier orbitals can satisfy the total density and orbital-potential constraints. Consequently, discrepancies arise; for instance, the SIC-LDA band gap of the Ar crystal reported by Heaton, Harrison, and Lin differs from that of Stengel and Spaldin as much as 0.7 eV \cite{cite-SIC-25-wannier,cite-SIC-wannier-Si}. Large uncertainties are reported for the band gaps of bulk Silicon \cite{cite-SIC-wannier-Si,cite-SIC-max-wannier} and transition-metal mono-oxides (MnO, FeO, CoO, and NiO)~\cite{cite-SIC-transition-1,cite-SIC-transition-2}. The SIC orbitals of those results were all filtered by satisfying the constraints of total density and orbital potential but using different localization definitions for the Wannier orbitals.  

More recently, in 2020, Shinde \etal introduced the use of conceptually related FLOs for filtering out the Wannier orbitals (WFLO) \cite{cite-SIC-fermi-solid}. As a result, the WFLO-SIC approach identifies proper SIC orbitals for solids, leading to a significant improvement in the band gap of the semi-local PBE functional \cite{cite-pbe}. 
Additionally, alternative localization strategies have an enormous potential to establish the SIC orbitals \cite{cite-SIC-complex-multiple, cite-SIC-max-wannier}, including the maximally localized Wanner orbitals (MLWOs) \cite{cite-MLWO} or the Edmiston-Ruedenberg (E-R) type orbitals~\cite{cite-localized-method-2}. The E-R orbitals are similar to MLWOs in molecular systems~\cite{cite-E-R-pi} and are essentially identical to MLWOs in extended periodic systems~\cite{cite-E-R-solid}. However, these works evaluate the SIC-DFA total energies in a non-self-consistent manner. To our knowledge, there is no report on the self-consistent SIC-DFA implementation that explicitly incorporates MLWOs or E-R localization.

Besides the SIC problems discussed above, there is another potential source of inaccuracy in most of the discussed PZ-SIC studies,~i.e., the usage of semi-local DFAs at the level of LDA or GGA~\cite{cite-SIC-intro, cite-SIC-9, cite-SIC-complex-molecule,cite-SIC-20-FullFermi, 
cite-SIC-21-Fermi-C60,cite-SIC-25-wannier, cite-SIC-25-wannier}. Since the SIC is inherently linked to the employed DFA via Eq.~(\ref{eq:SIC}), it is yet unclear to which extent the observed trends are related to the SIC or to the DFA. This is further substantiated by the fact that the combination of PZ-SIC and the state-of-the-art meta-GGA SCAN has shown some promising results~\cite{cite-SIC-18-Fermi-Theory,cite-meta-prescan,cite-SIC-SCAN,cite-SIC-water}. Consequently, there is an urgent need for practical, self-consistent SIC implementations compatible with the SCAN functional for both molecules and solids.

In this work, we overcome these challenges by building upon the \textit{total-density constraint} of Perdew and Zunger and the \textit{orbital potential constraint} of Pederson, Heaton, and Lin. We introduce a novel third constraint, the \textit{orbital density-potential constraint}, to PZ-SIC equations.
Inspired by the E-R orbitals~\cite{cite-localized-method-2}, our proposed constraint aims to maximize spatial localization of the SIC orbitals under the \textit{orbital potential constraint}. 
As demonstrated in the subsequent sections, this approach facilitates efficient convergence of the PZ-SIC equations for both molecular and extended systems. 

We demonstrate the efficacy of our proposed methodology through its implementation in the all-electron numeric atom-center orbital (NAO) package \textit{FHI-aims}~\cite{cite-aims-1, cite-aims-2}. 
Besides applying our PZ-SIC approach for semi-local DFAs such as LDA and GGA, 
we primarily focus on PZ-SIC calculations with the meta-GGA SCAN functional (SIC-SCAN). For the latter, we observe significant improvements in the highest occupied orbital energies when compared to the ionization potential of atoms from H to Ar, as well as for 18 molecules taken from reference~\cite{cite-GW100}. Furthermore, the band gaps of various solids, including Si, MnO, diamond-C, MgO, and LiF, are calculated using SIC-SCAN, yielding comparable results to $GW$ calculations. The combination of PZ-SIC and SCAN offers enhanced accuracy in electronic structure calculations while maintaining a moderate computational burden, particularly for periodic solids.

The paper is organized as follows: Section \ref{sec-theory} provides a derivation of the standard PZ-SIC formula and introduces the three constraints used to solve the PZ-SIC equations, both for finite and periodic systems. 
In particular, we analyze and discuss the physical motivation behind the \textit{orbital density-potential constraint} by using the carbon atom, the Helium dimer (He$_2$), the Helium crystal, and the periodic polyethylene chain. In section \ref{sec-results}, we assess the performance of PZ-SIC in conjunction with SCAN by examining various metrics such as SIC energy, ionization potential, band gap, and cohesive energy for a group of molecules and solids. Conclusions and outlook are presented in section \ref{sec-conclusion}.

\section{Theory} \label{sec-theory}
With the PZ-SIC concept (Eq.~(\ref{eq:SIC})), the total energy of a self-interaction corrected DFA, termed SIC-DFA, is written as~\cite{cite-SIC-4-manybody}
\begin{equation}
  \begin{aligned} \label{eq:SIC-Energy}
    & E^{\text{SIC-DFA}}[n, \{ n_i \}] :=  E^{\text{DFA}}[n] + E^{\text{PZ-SIC}}[\{n_{i}\}] \quad. \\
          \end{aligned}
\end{equation}
Here $E^{\text{DFA}}[n]$ is the total energy of a given DFA in the standard KS framework~\cite{cite-GEA-2}:
\begin{equation} \label{eq:KS-Energy}
  \begin{aligned} 
	  E^{\text{DFA}} [n] := T_{\text{s}} [n] + E_{\text{ext}} [n] +E_{\text{H}}[n] +E^{\text{DFA}}_{\text{xc}} [n] \quad,
    \end{aligned}
\end{equation}
where $T_\text{s}$ is the kinetic energy of the KS non-interacting electrons. 
$E_{\text{ext}}[n]$ and $E_{\text{H}}[n]$ are the external potential energy and the Hartree energy, respectively,
both of which are explicit functionals of the electron density $n (\tb{r})$.
$E^{\text{DFA}}_{\text{xc}}[n]$ is the XC energy of the DFA.

In the standard KS-DFT scheme, the KS equations are obtained via a variational minimization of the DFA energy $E^{\text{DFA}}[n]$ (Eq.~(\ref{eq:KS-Energy})), whereby the conservation of the number of electrons $N_e$ has to be ensured. This is equivalent to enforcing orthonormalization on the canonical occupied KS orbitals $\{ \psi_{l} \}$~\cite{cite-SIC-4-manybody}, leading to the Euler equation
\begin{equation} \label{eq:Euler-0}
  \begin{aligned}
	  \delta &\left[ E^\text{DFA} - \sum_{ij}^{N_e} \epsilon_{ij} \left( \dotstate{\psi_i}{\psi_j} - 
	  \delta_{ij} \right) \right] = 0\quad.
  \end{aligned}
\end{equation}
Because the DFA energy $E^{\text{DFA}}[n]$ remains invariant under the unitary transformations of the canonical occupied KS orbitals $\{ \psi_{l} \}$, any orthonormalized sets of single-particle orbitals $\{ \phi_{i} \}$, derived through a unitary transformation from $\{ \psi_{l} \}$, satisfies Eq.~(\ref{eq:Euler-0}) and yields the minimal DFA energy $E^{\text{DFA}}[n]$. It is worth pointing out that while the unitary transformation preserves the total density
\begin{equation} \label{eq:total-density}
  \begin{aligned}
	  \sum_{i=1}^{N_e} \left|\phi_i (\tb{r})\right|^2=\sum_{i=1}^{N_e} \left|\psi_i (\tb{r})\right|^2 = n(\tb{r}) \quad,
  \end{aligned}
\end{equation}
it does not necessarily maintain the individual orbital densities $\{ n_i(\textbf{r}) \}$
\begin{equation} \label{eq:orbital-density}
  \begin{aligned}
	  n_i(\tb{r}) = \left|\phi_i (\tb{r})\right|^2\neq \left|\psi_i (\tb{r})\right|^2 \quad.
  \end{aligned}
\end{equation}

The variational minimization of the SIC-DFA energy $E^{\text{SIC-DFA}}$ (Eq.~(\ref{eq:SIC-Energy})) is more complicated because the PZ-SIC contribution $E^{\text{PZ-SIC}}[\{n_{i}\}]$---the second part of Eq.~(\ref{eq:SIC-Energy})---is not invariant under unitary transformations of the occupied KS orbitals, $\{ \psi_{l} \}$. Consequently, there exists an alternative orthonormalized set of single-electron orbitals $\{ \phi_{i} \}$, which preserves the total density (Eq.~(\ref{eq:total-density})). This set can further minimize the PZ-SIC contribution and thus the overall SIC-DFA energy. These orbitals, $\{ \phi_{i} \}$, referred to as SIC orbitals, are derived from unitary transformations of the KS orbitals $\{ \psi_{l} \}$, which are not necessarily identical to the KS orbitals~\cite{cite-SIC-off-diagonal}.
Therefore, the effort to minimize $E^{\text{SIC-DFA}}$ (Eq.~(\ref{eq:SIC-Energy})) necessitates an additional orthonormalization constraint on the occupied SIC orbitals  $\{ \phi_{i} \}$~\cite{cite-SIC-complex-multiple,cite-SIC-8-lsda}, leading to the generalized Euler equation
\begin{equation} \label{eq:Euler-1}
  \begin{aligned}
	  \delta &\left[ E^\text{SIC-DFA} - \sum_{ij}^{N_e} \epsilon_{ij} \left( \dotstate{\psi_i}{\psi_j} - 
	  \delta_{ij} \right) \right.\\
	  & \left. - \sum_{ab}^{N_e} \lambda_{ab} \left( \dotstate{\phi_a}{\phi_b} - \delta_{ab} \right)\right] = 0\quad.
  \end{aligned}
\end{equation}

\subsection{PZ-SIC One-Electron Equations}
Unfortunately, the fact that two different sets of orbitals, i.e., $\{\phi_i\}$ and $\{\psi_i\}$, enter Eq.~(\ref{eq:Euler-1}) and that these orbitals are inherently coupled complicates the solution of this equation. In practice, the problem is approached by solving two distinct problems: one to determine the KS orbitals $\{ \psi_{l} \}$ and one for the SIC orbitals $\{ \phi_{i} \}$. To ensure self-consistency, these two sets of equations are solved iteratively. 

The PZ-SIC one-electron equations for the KS non-interacting systems are 
\begin{equation} \label{eq:SIC-eigen}
  \begin{aligned}
  \hat{h}^{\text{SIC-DFA}} \psi_{l} = 
       \epsilon_l \psi_{l} \quad.
  \end{aligned}
\end{equation}
Here, $\hat{h}^{\text{SIC-DFA}}$ is the KS non-interacting Hamiltonian for the SIC-DFA
\begin{equation} \label{eq:SIC-operator}
  \begin{aligned}
  \hat{h}^{\text{SIC-DFA}} =
  \hat{t}_{\text{s}} + \hat{v}_{\text{ext}} + \hat{v}_{\text{H}} + \hat{v}_{\text{xc}} 
  + \hat{v}^{\text{SIC}} \quad,
  \end{aligned}
\end{equation}
which includes the kinetic energy operator $\hat{t}_{\text{s}}$, 
the external potential $\hat{v}_{\text{ext}}$, the Hartree potential $\hat{v}_{\text{H}}$, 
the XC potential $\hat{v}_{\text{xc}}$, and the SIC operator $\hat{v}^{\text{SIC}}$. 
Compared to the standard KS Hamiltonian, Eq.~(\ref{eq:SIC-operator}) additionally features the SIC operator
\begin{equation} \label{eq:SIC-potential}
  \begin{aligned}
  \hat{v}^{\text{SIC}} \psi_{l} & = \frac{\delta E^{\text{PZ-SIC}}[\{n_i\}]}{\delta n} \cdot \psi_{l} = \sum_i^{N_e} \hat{v}_i^{1e\text{SIC}} \rstate{\phi_i} \dotstate{\phi_i}{\psi_{l}} \quad. \\
    \end{aligned}
\end{equation}
Here, the one-electron SIC potential $\{ \hat{v}^{1e\tf{SIC}}_i \}$ in the SIC operator is the negative of summing the single-particle Hartree potentials $\hat{v}_{\tf{es}}[n_i]$ and the DFA-specific single-particle XC potentials $\hat{v}_{\tf{xc}}[n_i]$, evaluated for the SIC orbital densities $n_i(\tb{r})$:
\begin{equation}
  \begin{aligned}
    \hat{v}^{1e\tf{SIC}}_i = - (\hat{v}_{\tf{H}}[n_i] + \hat{v}_{\tf{xc}}[n_i]) \quad 
  \end{aligned}
\end{equation} 
It is important to note that the SIC operator $\hat{v}^{\text{SIC}}$ depends on the SIC orbitals $\{\phi_i\}$. 
Consequently, the computations of the SIC operator in this work occur within the framework of the generalized Kohn-Sham (gKS) theory~\cite{cite-gKS-1,cite-gKS,cite-gKS-2} in analogy to the gKS derivation of the meta-GGA SCAN potential~\cite{cite-scan-1}. 
(Please refer to Ref.~\cite{cite-SIC-4-manybody, cite-SIC-16, cite-SIC-intro} for more detailed derivations).
Given a proper initial guess for $\{\phi_i\}$, the KS orbitals $\{\psi_i\}$ can be obtained by solving the PZ-SIC equations (Eq.~(\ref{eq:SIC-eigen})). From this, the ground-state density is derived as Eq.~(\ref{eq:total-density}), which in turn yields the minimal SIC-DFA total energy~(Eq.~(\ref{eq:SIC-Energy})) corresponding to the initial guess $\{\phi_i\}$~\cite{cite-SIC-complex-multiple}. In practical scenarios, the inherent challenge of multiple solutions in the standard PZ-SIC scheme manifests as a sensitivity of the self-consistent SIC orbitals to this initial guess. This phenomenon is further exemplified by the cases of the Carbon atom, Helium dimer, and Helium crystal in Sec.~\ref{sec:constraint} and Sec.~\ref{sec:constraintPBC}.

\subsection{PZ-SIC Constraints}\label{sec:constraint}
In the pursuit of minimizing the SIC-DFA total energy using the self-consistent PZ-SIC scheme, it is imperative to iteratively refine the SIC orbitals $\{\phi_i\}$ based on the KS orbitals $\{\psi_l\}$ from the prior iteration (Eq.~(\ref{eq:SIC-eigen})). This updating process for the SIC orbitals must adhere to specific physical constraints.
\newline

\noindent\textbf{Constraint 1.} Total density constraint (TDC):\\
The SIC orbitals $\{\phi_i\}$ are enforced to produce the same density $n(\textbf{r})$ as the canonical KS orbitals $\{\psi_l\}$:
\begin{equation} \label{eq:sum-rule}
\begin{aligned}
\sum_i^{N_e} | \phi_i (\textbf{r}) |^2 = n(\textbf{r}) = \sum_l^{N_e} | \psi_l (\textbf{r}) |^2\quad.
\end{aligned}
\end{equation}
Technically speaking, this implies that the occupied SIC orbitals must be related to the occupied KS orbitals via a unitary transformation~\cite{cite-SIC-4-manybody}:
\begin{equation} \label{eq:transform-T}
  \begin{aligned}
    & \phi_i = \sum^{N_{\text{e}}}_l T_{il} \psi_l \quad, \\
    & \delta_{ml}  = \sum_i^{N_e} T^*_{im} T_{il} \quad.
  \end{aligned}
\end{equation}

Under this constraint and under the assumption that the KS orbitals $\{\psi_l\}$ and the total density $n(\textbf{r})$ from the preceding iteration remain unchanged, the generalized Euler equation~(Eq.~(\ref{eq:Euler-1})) yields the one-electron SIC potential equations:
\begin{equation} \label{eq:SIC-eigen2}
  \begin{aligned}
	  \hat{v}^{1e \text{SIC}}_i \phi_i  = \sum_a \lambda_{ai} \phi_a \quad,
  \end{aligned}
\end{equation}
which lead to the Pederson constraint that follows in Eq.~(\ref{eq:potential_constraint})
(Please refer to Ref.~\cite{cite-SIC-16} for more detailed derivations).
\newline

\noindent\textbf{Constraint 2.} Orbital potential constraint (OPC):\\
\begin{equation} \label{eq:potential_constraint}
  \begin{aligned}
    \avstate{\phi_m}{ \hat{v}_m^{1e\text{SIC}}}{\phi_n} 
    = \avstate{\phi_m}{ \hat{v}_n^{1e\text{SIC}}}{\phi_n} \quad.
  \end{aligned}
\end{equation}
For the KS orbitals $\{\psi_l\}$ from the preceding iteration, the minimal SIC-DFA total energy (Eq.~(\ref{eq:SIC-Energy})) is provided by the SIC orbitals that satisfies this orbital potential constraint. The resulting SIC orbitals are frequently (but not always) localized~\cite{cite-SIC-8-lsda}.

In summary, to find the minimal SIC-DFA energy via the variational minimization scheme, we must solve the PZ-SIC one-electron equations self-consistently under Constraints 1 and 2 (TDC and OPC, Eqs.~(\ref{eq:sum-rule}) and (\ref{eq:potential_constraint})). The well-documented challenge of multiple solutions arises from the fact that the $E^{\tf{SIC-DFA}}$ functional (Eq.~(\ref{eq:SIC-Energy})) possesses many local minima. As a result, PZ-SIC outcomes are highly sensitive to the initial guess of SIC orbitals~\cite{cite-SIC-complex-multiple}.

As discussed in the introduction, the benefits of localization procedures have been established within the framework of PZ-SIC~\cite{cite-SIC-4-manybody}. This is a logical outcome since more localized densities generally result in stronger self-interaction, leading to a most significant reduction in the total energy within the SIC-DFA scheme. Moreover, localization procedures, as exemplified by FLO, can reduce the number of unitary transformations (Eq.~(\ref{eq:transform-T})), thereby facilitating the self-consistent PZ-SIC convergence. Inspired by these localized methods, we propose the construction of SIC orbitals by self-consistently enforcing additional strong localization. 
Note that different procedures and approaches exist for this purpose~\cite{cite-SIC-8-lsda,cite-SIC-max-wannier,cite-SIC-complex-multiple}. In this work, we choose the Edmiston-Ruedenberg (E-R) restriction~\cite{cite-localized-method-2}, which, by definition, maximizes the integrals over the squared densities of the individual SIC orbitals \begin{equation}
  \begin{aligned} \label{eq:max-squared}
	  \max \int d\textbf{r} \sum_i n^2_i(\textbf{r}) \Leftrightarrow \min \int d\textbf{r} \sum_{i \neq j} n_i(\textbf{r})n_j(\textbf{r}) \quad,
  \end{aligned} 
\end{equation}
which is equivalent to minimizing the overlap of different SIC orbital densities~\cite{cite-localized-method-3}, and thus maximizing the sum of the single-particle Hartree energy $\sum_{i}^{N_e}E_{\text{H}}[n_i]$~\cite{cite-localized-method-2}. For this reason, Perdew and Zunger suggested in the original PZ-SIC paper that E-R localized orbitals are an appropriate choice for the PZ-SIC framework as also confirmed by Pederson and Lin~\cite{cite-SIC-8-lsda}.

In order to incorporate the E-R restriction (Eq.~(\ref{eq:max-squared})) as a constraint into the self-consistent PZ-SIC framework, we follow the spirit and the derivation of the OPC to obtain the self-consistently localized SIC Euler equation
\begin{equation} \label{eq:slSIC-euler}
  \begin{aligned}
    \delta \left[ \int d^3\tb{r} \sum_i n^2_i(\tb{r}) - \eta (\delta E^\tf{SIC-DFA}) \right]  =0 \quad .
  \end{aligned}
\end{equation}
This variation leads to a new constraint, see the following Eq.~(\ref{eq:den-constraint}).
\newline

\noindent\textbf{Constraint 3.} Orbital density-potential constraint (ODPC):\\ 
\begin{equation} \label{eq:den-constraint}
\begin{aligned}
  \avstate{\phi_m}{ n_m  \hat{v}_n^{1e\tf{SIC}} }{\phi_n} = \avstate{\phi_m}{ \hat{v}_m^{1e\tf{SIC}} n_n }{\phi_n} \quad,
\end{aligned}
\end{equation}
which can be formulated as a set of localization equations (Please see Supplementary Sec. I for a detailed derivation of these relationships). The herein-introduced constraint based on the E-R restriction imposes an even stronger emphasis on localization than the orbital-potential constraint. As a consequence, it helps to steer the solution of the PZ-SIC equations towards the correct global minimum $E^\tf{SIC-DFA}$, especially when multiple solutions associated with nonlocal minima are possible if only the two first constraints are enforced.
First, we explain and illustrate this for two simple systems, He$_2$ and C.
Below, we will then further substantiate this by extended benchmarks for solids, see Sec.~\ref{sec:constraintPBC}.

\begin{figure}[htbp] 
\vspace{0.0in}
\begin{center}
\includegraphics[width=3.4in,angle=0]{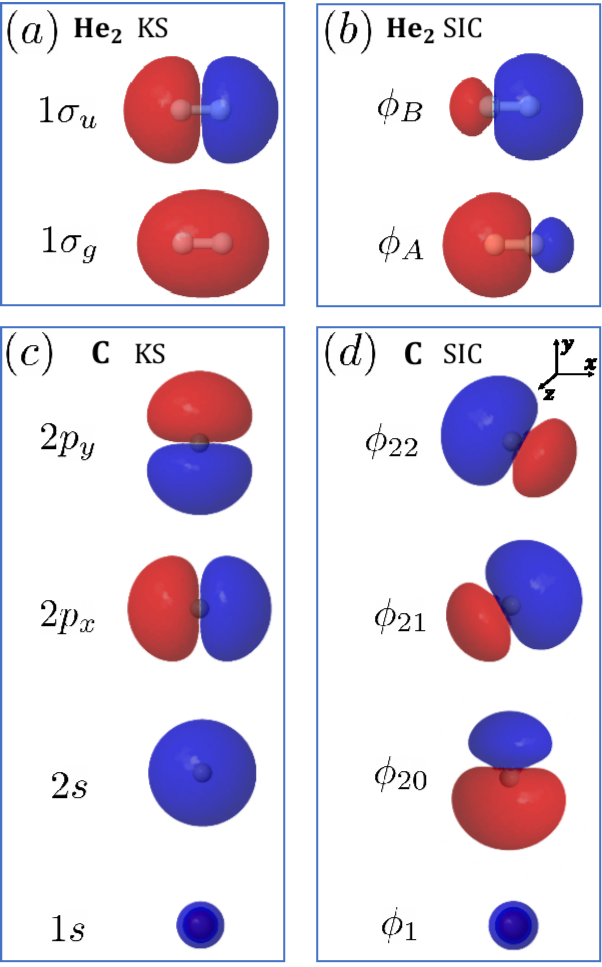}
\end{center}
\vspace{-0.2in}
\caption{Depictions of the KS orbital densities for the He$_2$ molecule in (a) and their corresponding SIC orbital densities in (b).
For the C atom, the KS orbital densities are presented in (c), and the SIC densities in (d).
The \text{(SIC-)SCAN} calculations in \ti{FHI-aims} were used to assess both sets of orbitals, KS and SIC, employing \ti{tight} \ti{tier}-1 NAO basis sets and default output settings. 
The results were visualized using the open-source tool \ti{Jmol}~\cite{jmol} with its predefined settings.
}
\label{Fig-LO}
\vspace{0.0in}
\end{figure}

As a first example, we consider the closed-shell dihelium molecule He$_2$ with a bond length of 1.1 \AA.
Here, $1\sigma_g$ and $1\sigma_u$ are the fully occupied KS orbitals featuring two electrons each. 
These occupied KS orbitals are delocalized over the atoms, cf.~ Fig.~\ref{Fig-LO}(a). 
As derived in Supplementary Sec. II, there are two sets of orbitals that fulfill both TDC and OPC. These are the canonical KS orbitals $\{1\sigma_{g;u}\}$ and the more localized orbitals
 $\{\phi_{A;B}\}$, see Fig.~\ref{Fig-LO}(b), that can be obtained via the unitary transformation:
\begin{equation}
  \begin{aligned}
    \phi_{A;B} = \frac{1}{\sqrt{2}} [1\sigma_g \pm 1\sigma_u ] \quad.
  \end{aligned}
\end{equation}
More specifically, Constraint 1 (TDC, Eq.~(\ref{eq:sum-rule})) is fulfilled via
\begin{equation}
\begin{aligned}
	n(\textbf{r})   &= 2|1\sigma_{g}(\textbf{r})|^2 + 2| 1\sigma_{u}(\textbf{r}) |^2 \\
	                &= 2|\phi_A(\textbf{r})|^2 + 2| \phi_B(\textbf{r})|^2 \quad,
\end{aligned}
\end{equation}
and Constraint 2 (OPC, Eq.~(\ref{eq:potential_constraint})) via:
\begin{equation}
\begin{aligned}
	& \avstate{1\sigma_{g}}{ \hat{v}^{1e\text{SIC}}_{g}  }{1\sigma_{u}} = \avstate{1\sigma_{g}}{  \hat{v}^{1e\text{SIC}}_{u} }{1\sigma_{u}} \quad, \\
  & \avstate{\phi_A}{ \hat{v}^{1e\text{SIC}}_{A} }{\phi_B} = \avstate{\phi_A}{ \hat{v}^{1e\text{SIC}}_{B} }{\phi_B} \quad.
\end{aligned}
\end{equation}
This means that the SIC orbitals cannot be uniquely determined via Constraints 1 and 2 (TDC and OPC), even in this relatively simple case, in which $\{\phi_{A;B}\}$ are the obviously better, more localized choice.

This hurdle is overcome by Constraint 3 (ODPC, Eq.~(\ref{eq:den-constraint})), which is only fulfilled by the set of $\{\phi_{A;B}\}$, but not by $\{ 1\sigma_{g;u} \}$:
\begin{equation}
  \begin{aligned}
  & \avstate{1\sigma_{g}}{ n_{g} \hat{v}^{1e\text{SIC}}_{u}  }{1\sigma_{u}} \neq \avstate{1\sigma_{g}}{ \hat{v}^{1e\text{SIC}}_{g} n_{u} }{ 1\sigma_{u} } \quad, \\
  & \avstate{\phi_A}{ n_{A} \hat{v}^{1e\text{SIC}}_{B}  }{\phi_B} = \avstate{\phi_A}{ \hat{v}^{1e\text{SIC}}_{A}  n_{B} }{\phi_B} \quad .
  \end{aligned}
\end{equation}
As expected, the $\{ \phi_{A;B} \}$ orbitals then also yield a SIC-SCAN total energy that is about 5.29 eV lower than that of the $\{ 1\sigma_{g;u} \}$ orbitals \footnote{ The SIC-SCAN calculations for He$_2$ were carried out with FHI-aims using the \ti{tight} \ti{tier}-1 NAO basis set.}.
This is in line with the original intention of the constraint to enforce localization and showcases how ODPC is a useful and often necessary complement to PZ-SIC theory.  

Second, we showcase a similar but less trivial example, the SIC-SCAN total energy calculations for the carbon atom with the electronic ground state configuration $1s2s (2p)^2$. The atomic orbitals $\{1s, 2s, 2p_x, 2p_y\}$ shown in Fig.~\ref{Fig-LO}(c) satisfy both TDC and OPC \cite{cite-SIC-8-lsda} but not ODPC. However, all three constraints can be simultaneously met by the following set of orbitals obtained from a unitary transformation of the atomic orbitals:
\begin{equation} \label{eq:C-orbital}
  \begin{aligned}
    & \phi_{1} = 1s \\
    & \phi_{20} = \frac{1}{\sqrt{3}} 2s - \frac{\sqrt{2}}{\sqrt{3}} 2p_x \\
    & \phi_{21} = \frac{1}{\sqrt{3}} 2s + \frac{1}{\sqrt{6}} 2p_x + \frac{1}{\sqrt{2}} 2p_y \\
    & \phi_{22} = \frac{1}{\sqrt{3}} 2s + \frac{1}{\sqrt{6}} 2p_x - \frac{1}{\sqrt{2}} 2p_y \quad . \\
  \end{aligned}
\end{equation}
As shown in Fig.~\ref{Fig-LO}, 
these hybridized orbital densities ($|\phi_{20}|^2$ and $|\phi_{21}|^2$) break the spherical harmonic symmetry and are thus more localized, 
i.e., they exhibit a smaller overlap compared to the atomic orbital densities ($|2s|^2$ and $|2p_x|^2$). 
Accordingly, these hybridized orbitals defined in Eq.~(\ref{eq:C-orbital}) result in a SIC-SCAN energy that is about 0.86 eV lower than that obtained using atomic orbitals.

\subsection{Periodic Systems} \label{sec:constraintPBC}
In periodic boundary conditions (PBCs), the PZ-SIC one-electron equations have to be solved for 
multiple $\textbf{k}$-points in the first Brillouin zone
\begin{equation} \label{eq:SIC-eigen-k}
  \begin{aligned}
	  & \hat{h}^{\text{SIC-DFA}} \psi_{l,\textbf{k}} = \epsilon_{l,\textbf{k}} \psi_{l,\textbf{k}} \quad.
  \end{aligned}
\end{equation} 
The KS orbitals $\{\psi_{l,\textbf{k}}(\textbf{r})\}$ are the generalized Bloch orbitals located in the first Brillouin zone.
Due to the localized nature of the SIC orbitals, Heaton, Harrison, and Lin proposed that it is more convenient to express the SIC orbitals $\phi_{i, \textbf{I}}(\textbf{r})$ 
in terms of Wannier orbitals $\{\Psi_{l, \textbf{L}}(\textbf{r})\}$ \cite{cite-SIC-25-wannier}. Accordingly,  the SIC orbital $\phi_{i, \textbf{I}}(\textbf{r})$ centered in the $\textbf{I}$th unit cell is expressed as:
\begin{equation} \label{eq:transform-T-w}
  \begin{aligned}
	  & \phi_{i, \textbf{I}}(\textbf{r}) = \sum_{l}^{N_e}\sum_{\textbf{L}}^{N_L} T_{i\textbf{I},l\textbf{L}} \Psi_{l, \textbf{L}}(\textbf{r}) \quad. \\
          \end{aligned}
\end{equation}
Here, $N_e$ is the number of electrons in a unit cell, and $N_L$ is the unit cell number (also the \textbf{k}-point number) in the supercell. 
The Wannier orbitals $\{\Psi_{l, \textbf{L}}(\textbf{r})\}$ can be generated via a Fourier transformation of the KS orbitals $\{\psi_{l,\textbf{k}}(\textbf{r})\}$ and thus span the same space as the KS orbitals,
\begin{equation}
  \begin{aligned}
     \Psi_{l, \textbf{L}}(\textbf{r})  &= \Psi_{l} (\textbf{r} - \textbf{R}_{\textbf{L}} ) \\
	 & =  \frac{1}{\sqrt{N_L}}\sum_{\textbf{k}}^{N_L} \exp[-i \textbf{k} \cdot \textbf{R}_{\textbf{L}}] \psi_{l, \textbf{k}}(\textbf{r}) \quad.
  \end{aligned}
\end{equation}
Here, $\{\textbf{R}_{\textbf{L}}\}$ are the lattice vectors, with the subscript $\textbf{L}=[L_1,L_2,L_3]$ being the index of a given lattice vector.

In such a Wannier representation, TDC reads
\begin{equation} \label{eq:pbc-sum-rule}
  \begin{aligned}
	  \frac{1}{N_L}\sum_{ i \textbf{I}}^{N_e N_L}  | \phi_{i, \textbf{I}} (\textbf{r}) |^2 = n(\textbf{r}) = \frac{1}{N_L}\sum_{l,\textbf{k}}^{N_e N_L} | \psi_{l,\textbf{k}} (\textbf{r}) |^2 \quad.
  \end{aligned}
\end{equation}
Similarly, OPC and ODPC are expressed as
\begin{equation} \label{eq:pbc-constraint2}
  \begin{aligned}
    & \avstate{\phi_{m, \tb{M}}}{ \hat{v}_{m, \tb{M}}^{1e\tf{SIC}}}{\phi_{n, \tb{N}}} 
    = \avstate{\phi_{m, \tb{M}}}{ \hat{v}_{n, \tb{N}}^{1e\tf{SIC}}}{\phi_{n, \tb{N}}}     
  \end{aligned}
\end{equation}
and
\begin{equation} \label{eq:pbc-constraint3}
  \begin{aligned}
    & \avstate{\phi_{m, \tb{M}}}{ n_{m, \tb{M}} \hat{v}_{n, \tb{N}}^{1e\tf{SIC}}}{\phi_{n, \tb{N}}} 
    =  \avstate{\phi_{m, \tb{M}}}{ \hat{v}_{m, \tb{M}}^{1e\tf{SIC}} n_{n, \tb{N}} }{\phi_{n, \tb{N}}} \quad .
  \end{aligned}
\end{equation}
In this case, $\hat{v}^{1e\text{SIC}}_{i, \textbf{I}} = \delta E^{\text{PZ-SIC}} [\{n_{i,\textbf{I}}\}] / \delta n_{i,\textbf{I}} $ is the one-electron SIC potential under PBCs, and $n_{i,\textbf{I}} = |\phi_{i, \textbf{I}}|^2 $ is the associated SIC orbital density
(Please refer to Ref.~\cite{cite-SIC-25-wannier, cite-SIC-fermi-solid} and Supplementary Sec. I for more detailed derivations).

\begin{figure*}[htbp] 
  \vspace{0.0in}
  \begin{center}
    \includegraphics[width=6.9in,angle=0]{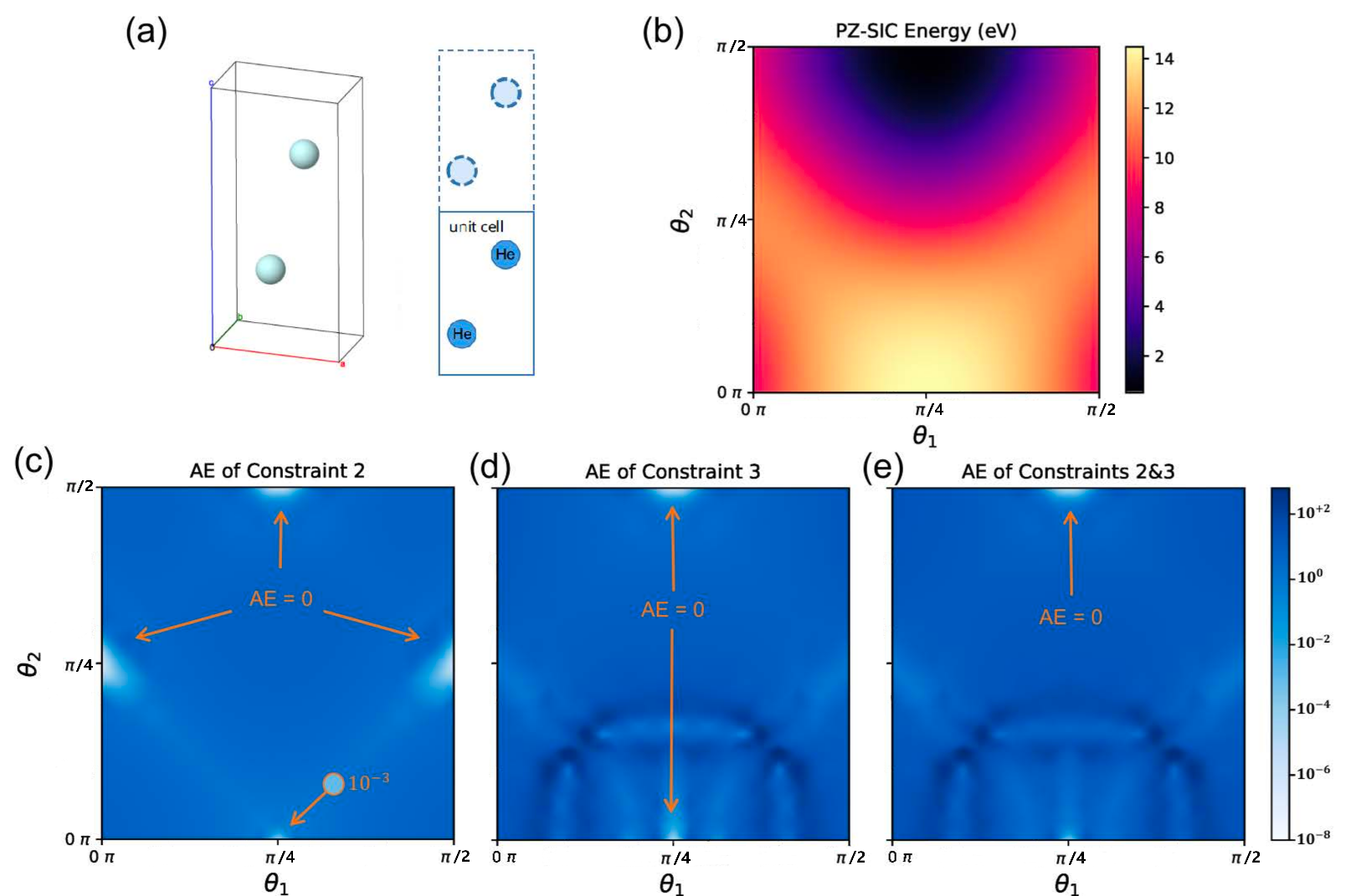}
  \end{center}
  \vspace{-0.2in}
  \caption{
    (a) Schematic representation of the hcp Helium crystal's unit cell on the left; while the depiction of two unit cells incorporating periodic boundary conditions on the right. Geometry was taken from Ref.~\cite{cite-He-geometry}.
            (b) The energy profile of $E^\tf{PZ-SIC}$ (Eq.~(\ref{eq:SIC})) with respect to the rotation angles ($\theta_1, \theta_2$).
    (c) The absolute errors (AE) defined in Eq.~(\ref{eq:AE}) pertaining to OPC (Constraint 2) and (d) ODPC (Constraint 3), respectively.
    (e) Combined AE of both OPC and ODPC for all possible orbitals in the $1\times1\times2$ Helium unit cell.
    These orbitals are unitarily transformed from the $1\Sigma_{g, \tb{I}}, 1\Sigma_{u, \tb{I}}$ orbitals using rotation angles $\theta_1$ and $\theta_2$.
    In the subfigures (c-e), the regions in white indicate the solutions for OPC, ODPC, and their combinations, respectively.
    Both SCAN and SIC-SCAN calculations in \ti{FHI-aims} employed PBCs and \ti{tight} \ti{tier}-1 NAO basis sets.
        }
  \label{Fig-He}
  \vspace{0.0in}
\end{figure*}

As mentioned in the introduction, the SIC multiple-solution problem is known to be more serious for solids~\cite{cite-SIC-25-wannier,cite-SIC-transition-1,cite-SIC-transition-2,cite-SIC-5-atom-orbital,cite-SIC-wannier-Si}. To illustrate this and to showcase how the introduction of ODPC helps to alleviate this issue, we here discuss crystal He, a prototypical case with a multiple-solution problem in standard SIC approaches. 
As there are two He atoms in each unit cell, see Fig.~\ref{Fig-He}(a),  
the occupied KS orbitals in the Wannier representation can be marked as $1\Sigma_{g, \tb{I}}$ and $1\Sigma_{u, \tb{I}}$. 
Naturally, these orbitals have the same shapes as the occupied molecular orbitals in the closed-shell dihelium He$_2$ molecule, see Fig.~\ref{Fig-LO}-(a). All orbitals $\{(\Phi_{1, \tb{I}}, \Phi_{2, \tb{I}})\}$ that satisfy TDC can be expressed via unitary transformations of the Wannier KS orbitals ($1\Sigma_{g, \tb{I}}, 1\Sigma_{u, \tb{I}}$). In this particular simple case, two distinct angles
$\theta_{I =1,2} \in [0, \pi/2]$ can be used to characterize this unitary transformation, as derived in detail in Supplementary Sec. III.
Accordingly, the original Wannier KS orbitals ($1\Sigma_{g, \tb{I}}, 1\Sigma_{u, \tb{I}}$) are obtained by choosing the angles to be $(\theta_{1} = 0, \theta_{2} = \pi/4)$.

To visualize the influence of the actual choice of  $\{(\Phi_{1, \tb{I}}, \Phi_{2, \tb{I}})\}$, we here show $E^{\text{PZ-SIC}}$ as an energy surface as function of $\theta_{I =1,2}$ in Fig.~\ref{Fig-He}-(b). The global $E^{\text{PZ-SIC}}$ minimum of 0.32 eV is obtained for the SIC orbitals $\Phi_{1;2, \tb{I}}$ with 
$(\theta_1 = \pi/4, \theta_2 = \pi/2)$. Respective SIC orbitals at these angles give $\frac{1}{\sqrt{2}} [ 1\Sigma_{g, \tb{I}} \pm 1\Sigma_{u, \tb{I}} ]$, which, not too surprisingly, localize at He atoms and have similar shapes as in the case of the non-periodic, closed-shell He$_2$ molecule, see Fig.~\ref{Fig-LO}-(b). 
At variance with Fig.~\ref{Fig-He}-(b), which is produced using all three constraints, Fig.~\ref{Fig-He}-(c) and Fig.~\ref{Fig-He}-(d) show what happens if only TDC and OPC viz. TDC and ODPC are used, respectively.
Additionally, Fig.~\ref{Fig-He}-(e) shows, once more, the results of using all three constraints. 
To better visualize the differences in these cases, 
we here locate the solutions of constraints in the individual plots by using the absolute error~(AE) of OPC or ODPC:
\begin{equation} \label{eq:AE}
  \begin{aligned}
    \text{AE} = \left\{
      \begin{aligned}
        & \text{max} \{ | \avstate{\phi_{m, \tb{M}}}{ \hat{v}_{m, \tb{M}}^{1e\tf{SIC}} -\hat{v}_{n, \tb{N}}^{1e\tf{SIC}} }{\phi_{n, \tb{N}}} | \} \\
        & \text{max} \{  \\
        & ~~~ | \avstate{\phi_{m, \tb{M}}}{ n_{m, \tb{M}} \hat{v}_{n, \tb{N}}^{1e\tf{SIC}} - \hat{v}_{m, \tb{M}}^{1e\tf{SIC}} n_{n, \tb{N}} }{\phi_{n, \tb{N}}} | \} \\
      \end{aligned}
    \right. \quad .
  \end{aligned}
\end{equation}
This reveals that the standard SIC method, which only accounts for TDC and OPC, is multi-valued 
since the three solutions at (0, $\pi/4$), ($\pi/2$, $\pi/4$), and ($\pi/4$, $\pi/2$) satisfy TDC and OPC, as shown in Fig.~\ref{Fig-He}-(c). Similarly,   
using only TDC and ODPC also yields multi-valued solutions at ($\pi/4$, 0.0) and ($\pi/4$, $\pi/2$), as shown in Fig.~\ref{Fig-He}-(d). 
The solution becomes uniquely single-valued only when all three constraints are employed, as shown in Fig.~\ref{Fig-He}-(e).
In this case,
 only one solution, i.e., ($\pi/4$, $\pi/2$), fulfills all constraints. This position also corresponds to the minimal SIC energy shown 
in Fig.~\ref{Fig-He}-(a).

\begin{figure}[htbp] 
  \begin{center}
    \includegraphics[width=3.4in,angle=0]{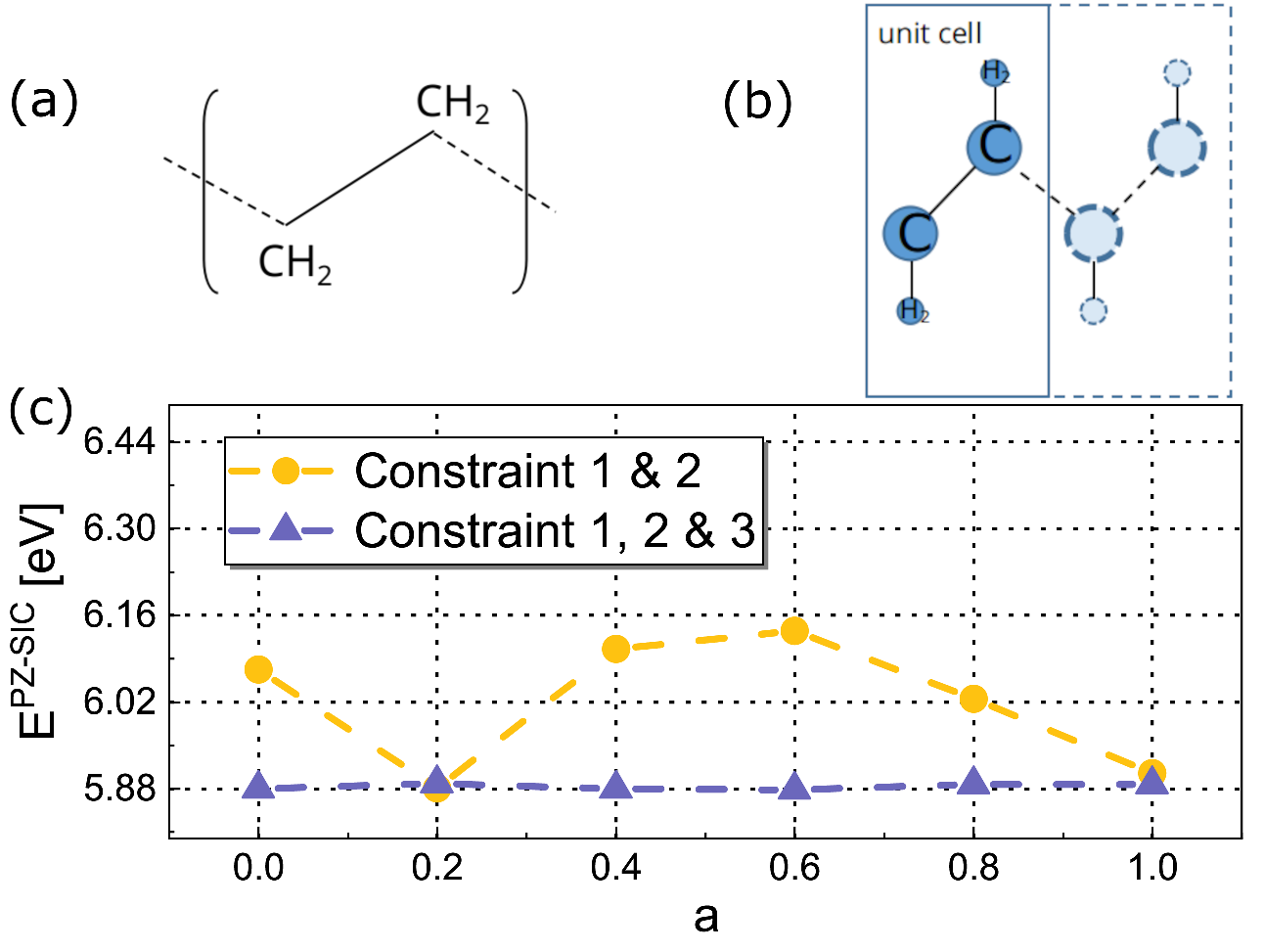}
  \end{center}
    \caption{ 
    (a) Primitive unit cell of the periodic polyethylene chain, $($C$_2$H$_4)_n$. 
    (b) Schematic view of the periodic polyethylene chain with two unit cells along the chain direction.
    Geometry was taken from Ref.~\cite{cite-CH2}.
    (c) PZ-SIC energies based on the SCAN functional (Eq.~(\ref{eq:SIC})). Different initial guesses of SIC orbitals were generated with respect to the mixing factor $a$ defined in Eq.~(\ref{eq:ini}).  
    All calculations were performed using \ti{tight} \ti{tier}-1 NAO basis sets in \ti{FHI-aims}.
    }
  \label{Fig-CH2-ini}
\end{figure}

As a last example, we turn our attention to the polyethylene chain, $($C$_2$H$_4)_n$. Its geometry is depicted in Fig.~\ref{Fig-CH2-ini}-(a) and (b). 
In such a complex system, the SIC orbitals cannot be exhaustively explored using a low-dimensional parametric model, unlike the simpler model systems previously discussed. Nonetheless, we here investigate the dependence of the SIC procedure by employing a set of initial guesses for the SIC orbitals, which are grounded in physical reasoning. These initial guesses are formulated by mixing the atomic orbitals $\varphi_{i} $ centered at the atomic position $\tb{R}^{ac}_{i}$ and the Wannier orbitals $ \Psi^{\tf{SCAN}}_{i, \tb{L}} $ obtained from SCAN calculations without including SIC:
\begin{equation} \label{eq:ini}
  \begin{aligned} 
    \phi^{\tf{init}}_{i, \tb{I}} = & \frac{1}{ A_{i}}  \left[ \right.  a  \varphi_{i} (\tb{r} - \tb{R}^{ac}_{i} -\tb{R}_{\tb{I}} )  \\
	&  + (1-a) \Psi^{\tf{SCAN}}_{i} (\tb{r} - \tb{R}_{\tb{I}} ) \left. \right] \quad.
  \end{aligned}
\end{equation}
Here, $A_{i}$ denotes a normalization factor, and $a$ is a mixing factor that ranges from $0$ to $1$. 
Accordingly, turning up $a$ from 0 to 1, the initial guess $\phi^{\tf{init}}_{i, \tb{I}}$ can be varied from having the pure atomic-orbital character to being a  Wannier orbital,~i.e.,~a KS molecular orbitals associated with the unit cell ${\tb{I}}$  
in real space. This allows tuning the localization of the initial guess, since, as discussed in the case of He$_2$, the atom orbitals that surround the atomic centers can be expected to be more localized than the Wannier orbitals \cite{cite-SIC-5-atom-orbital}. 
As shown in Fig.~\ref{Fig-CH2-ini}-(c), different solutions are found for different values of~$a$, with energies varying by about 0.3~eV at most, if only TDC and OPC are taken into account. This indicates that the problem has multiple local minima. Let us emphasize that even when starting from fully localized orbitals~$a=0$, TDC and OPC are not converging to the lowest energy solution. However, this doesn't apply when ODPC is additionally imposed because, in this case, all initial guesses converge to the same lowest-energy solution.

\section{Results} \label{sec-results}
We have implemented the aforementioned PZ-SIC method within the \ti{FHI-aims} package~\cite{cite-aims-1}. Detailed aspects of the implementation can be found in Supplementary Sec. IV. We here present benchmark results for a range of finite and periodic systems, including fundamental atoms, molecules, and bulk solids. It is worth noting that there are two distinct approaches for representing SIC orbitals in finite systems~\cite{cite-SIC-complex}.
One can either use a \ti{real}-valued unitary transformation matrix $T$ or a \ti{complex} one (Eq.~(\ref{eq:transform-T})). While both \ti{real} and \ti{complex} transformation matrices yield identical total densities, the resulting SIC energies can vary~\cite{cite-SIC-complex-noded}. 
This section will present results for both \ti{real} and \ti{complex} SIC orbitals when dealing with finite systems.
For periodic systems, where KS orbitals
are inherently \ti{complex}, we will exclusively consider \ti{complex} SIC orbitals.

\subsection{Finite Systems}

\subsubsection{Ionization Potential} \label{sec:IP}
The ionization potential (IP) measures the capability of an element to participate in chemical reactions that necessitate ion formation or electron transfer. IP is typically calculated via the total energy difference between the neutral molecule and the corresponding ion. Within the framework of (g)KS approach, as applied in specific DFAs, the IP is theoretically equivalent to the negative of the energy of the highest occupied molecular orbital (HOMO), as delineated in Ref.~\cite{cite-SIC-26-Yang-GSC, cite-PBEh}. However, this relationship often encounters deviations due to the errors, most notably the many-electron SIE. While gKS outperforms KS in orbital energy calculations~\cite{cite-scan-gKS-solid}, these errors still lead to significant discrepancies in the IP and predicted HOMO energies using (g)KS approaches~\cite{cite-PBEh}. In other words, this inconsistency affects the accuracy of IP estimations derived from HOMO energies. Consequently, the precision of calculated IPs based on HOMO energies remains a crucial metric for evaluating the effectiveness of various electronic-structure theory methodologies, including SIC~\cite{cite-GW100}.

\begin{figure}[!htbp] 
    \begin{center}
  \includegraphics[width=3.4in,angle=0]{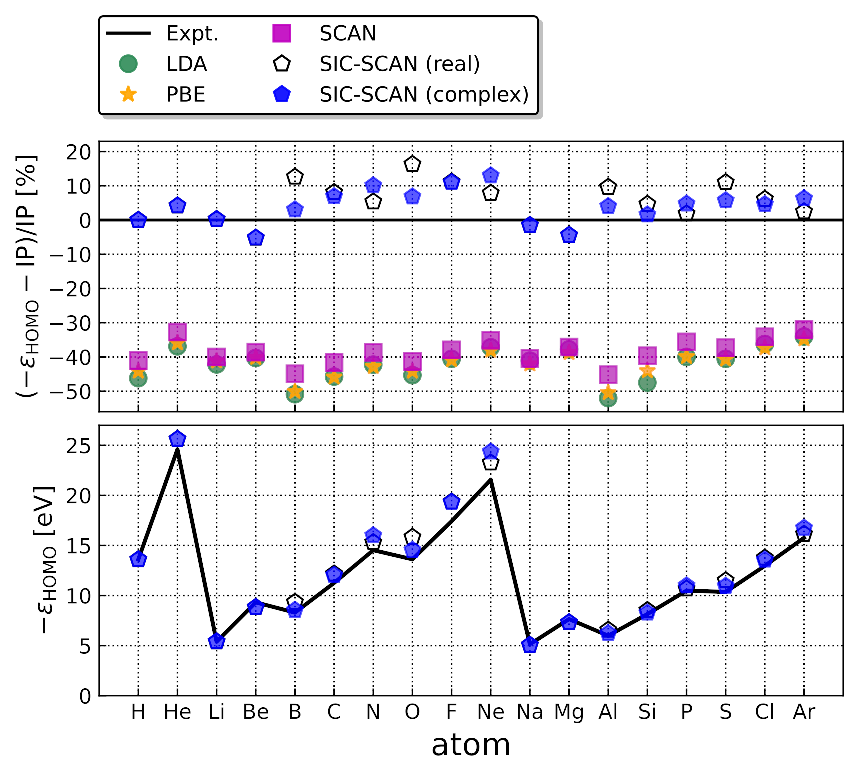}
  \end{center}
    \caption{ 
    Performance of various DFAs in describing the ionization potentials for the atoms ranging from H to Ar.
        SIC-SCAN calculations using \ti{real} and \ti{complex} SIC orbitals are marked as SIC-SCAN(real) and SIC-SCAN(complex), respectively.
    (top) Relative errors of $-\epsilon_{\tf{HOMO}}$ energies to IPs, while the $-\epsilon_{\tf{HOMO}}$ values are given in (bottom).
    All calculations were performed using \ti{tight} \ti{tier}-1 NAO basis sets.
    The experimental IPs were taken from the NIST database~\cite{cite:CCCBDB}.
  }
  \label{Fig-HOMO-atom}
  \end{figure}

The HOMO energies for atoms ranging from H to Ar, as calculated by LDA, PBE, SCAN, and SIC-SCAN, are presented in Fig.~\ref{Fig-HOMO-atom}, with experimental values~\cite{cite:CCCBDB} provided for reference. The pronounced SIE in LDA and PBE~\cite{cite-DFT-discontinuity} leads to a notable underestimation of the IPs. The performance of SCAN offers a marginal improvement. The mean absolute percentage errors (MAPE) are 42.03\% for LDA, 41.75\% for PBE, and 38.5\% for SCAN, substantiating that SCAN also suffers under SIEs. However, with SIC-SCAN, there is a notable improvement in the predictions. As shown in Fig.~\ref{Fig-HOMO-atom}-top, the MAPE is reduced to 5.02\% when using \ti{real} SIC orbitals and further drops to 3.95\% for \ti{complex} SIC orbitals.
This is in line with the findings in Ref.~\cite{cite-SIC-complex}, where the use of \ti{complex} SIC orbitals yielded superior results compared to their \ti{real} counterparts.

\begin{figure}[!htbp] 
    \begin{center}
  \includegraphics[width=3.4in,angle=0]{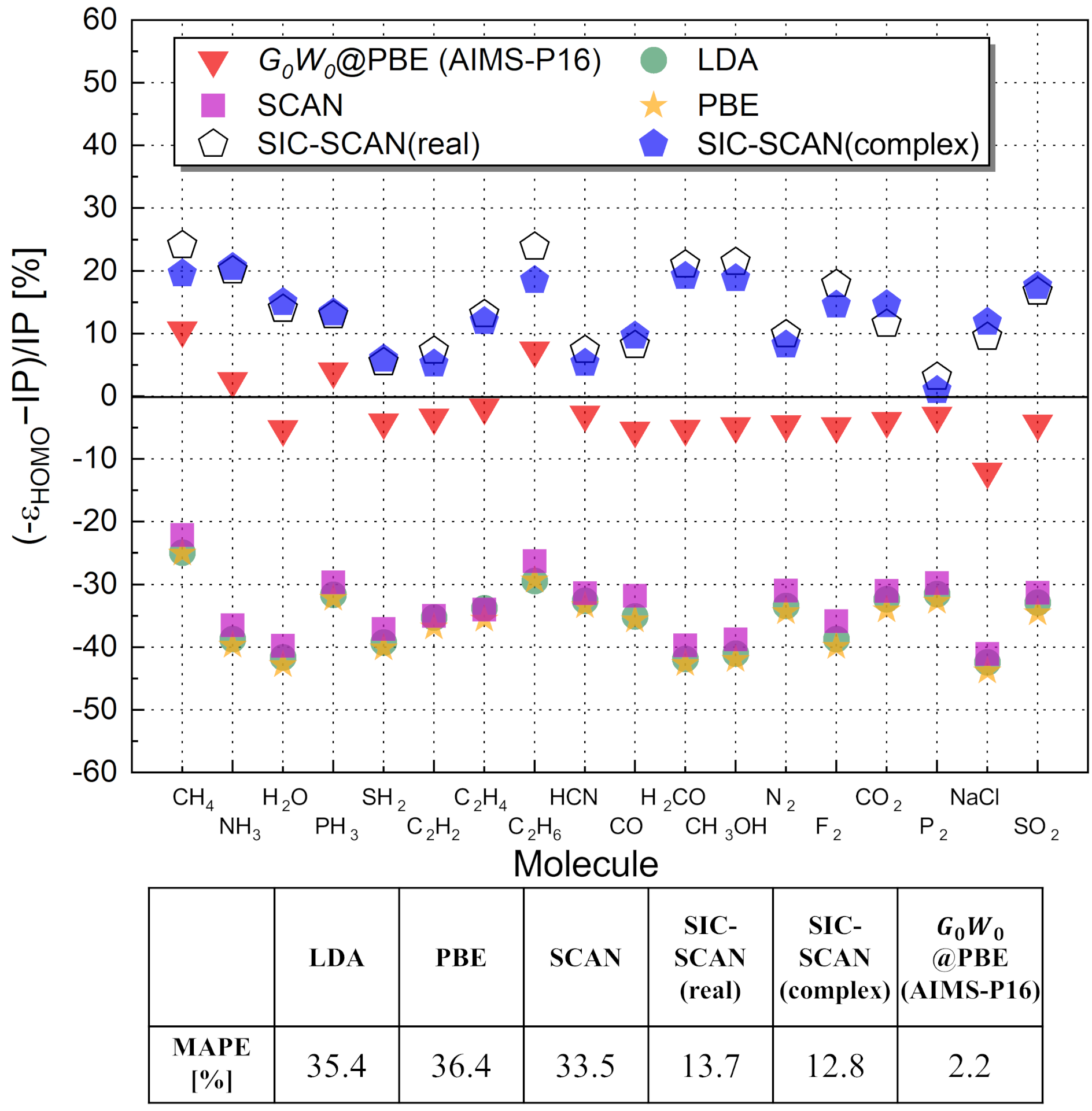}
  \end{center}
    \caption{ 
  (top) Performance of various DFAs in describing the ionization potentials of 18 molecules.
      The mean absolute percentage errors (MAPEs) of HOMO energies are given in (bottom).
  Calculations of LDA, PBE, SCAN, and SIC-SCAN methods were carried out in \ti{FHI-aims} using \ti{tight} \ti{tier}-1 NAO basis sets.
  $G_0W_0$@PBE results were taken from Ref.~\cite{cite-GW100}, which were also calculated in \ti{FHI-aims} but utilized def2-QZVP basis sets.
  The experimental vertical ionization energies, taken from the NIST database~\cite{cite:CCCBDB}, are reported for comparison.
  }
  \label{Fig-HOMO-mo}
  \end{figure}

Fig.~\ref{Fig-HOMO-mo} showcases the performance of various DFAs in predicting negative HOMO energies ($-\epsilon_\tf{HOMO}$) for a set of 18 molecules. 
As observed, LDA, PBE, and SCAN calculations tend to underestimate the $-\epsilon_\tf{HOMO}$ energies significantly, whereas the SIC approach considerably improves the results.
Specifically, all absolute percentage errors of SIC-SCAN(complex) 
are under 23\%, and its MAPE is $12.8\%$, which is substantially lower than the SCAN error ($33.5\%$). We note, however, that 
the PZ-SIC method  tends to slightly over-correct the SCAN calculations, resulting in the $-\epsilon_\tf{HOMO}$ energies consistently higher than 
the experimental IPs. To contextualize these results, $G_0W_0$@PBE, as reported in Ref.~\cite{cite-GW100}, 
offers IP predictions with a mere $3\%$ MAPE.

\subsubsection{Energy Curve and Broken Symmetry}
\begin{figure}[!htbp] 
    \begin{center}
            \includegraphics[width=3.4in, angle=0]{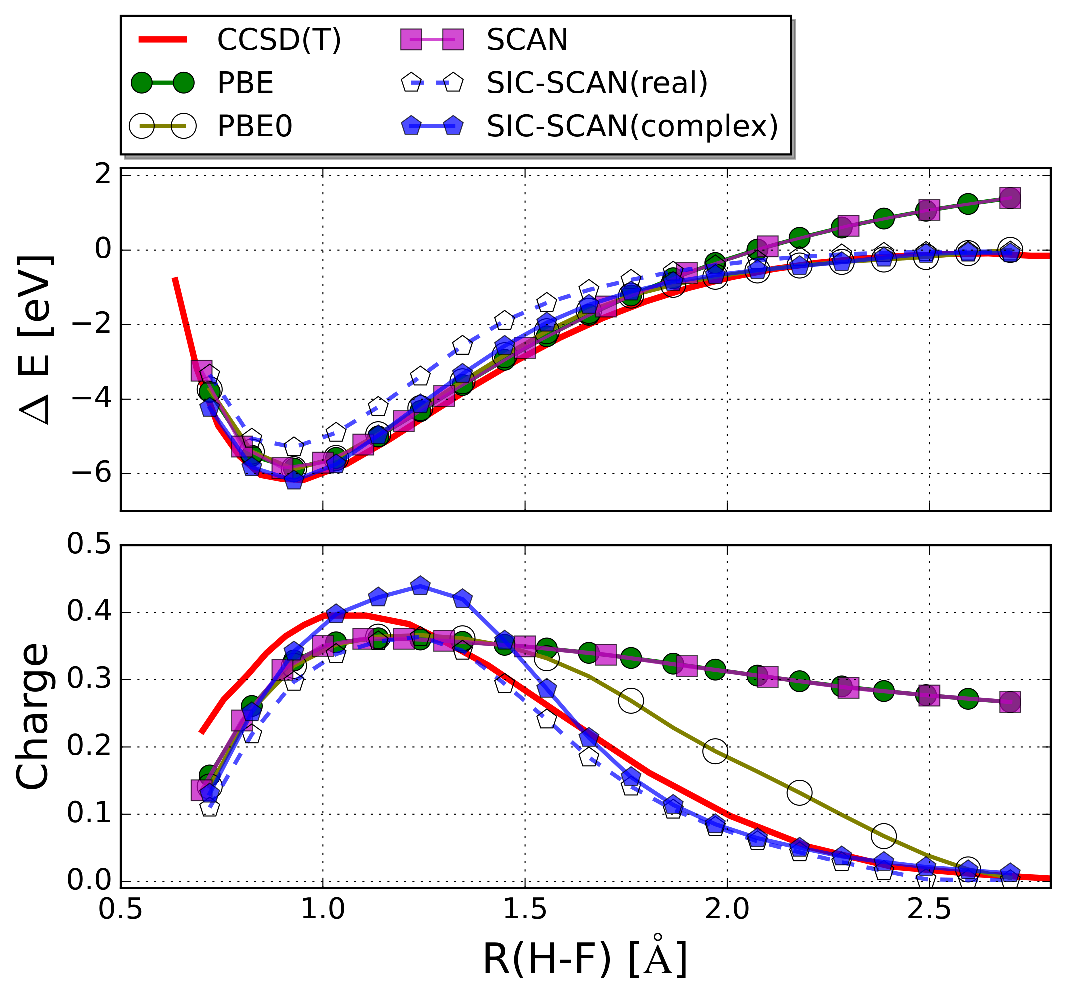}
  \end{center}
    \caption{ 
  (top) HF dissociation energy curves of various methods with the zero-energy level set to the total energy of isolated atoms/ions.
  (bottom) Mulliken charge analyses on the H atom along the dissociation.
  All calculations were performed in \ti{FHI-aims}. \ti{Tight} \ti{tier}-1 NAO basis sets were employed for PBE, PBE0, SCAN and SIC-SCAN incorporating \ti{real} and \ti{complex} SIC orbitals, while the CCSD(T) results utilize cc-pVTZ Gaussian-type basis sets.
  }
  \label{Fig-dissociation-AB}
  \end{figure}

As the SIE tends to delocalize electrons, semi-local DFAs typically struggle to describe charge transfer processes. A typical example is that semi-local DFAs tend to overly preserve the ground state symmetry of dissociating neutral heterodimers by producing fractionally charged fragments. 
For instance, SCAN incorrectly predicts a positive fractional charge on the H atom in the dissociation of the molecule H-F (Fig.~\ref{Fig-dissociation-AB}), hence giving too high energies in the dissociation limit. While Kim \etalcite~\cite{dc-improve} highlighted that a meticulous selection of orbital occupations can address this issue, the hybrid functional PBE0~\cite{cite:pbe0} outperforms both PBE and SCAN with self-consistent calculations. This superiority is attributed to incorporating a fraction (0.25) of Fock exchange in PBE0, which reduces the SIE in PBE. Nevertheless, PBE0 cannot correctly predict the change distribution at distances between 1.5-2.5 Å, as illustrated in Fig.~\ref{Fig-dissociation-AB}-bottom. In contrast, SIC reproduces comparable charge analyses of references and successfully maintains electron-neutral atoms in the dissociation limit.
Here, accurate results from coupled cluster theory with single, double, and perturbative triple excitations (CCSD(T)) are used as references. As shown, the energy curve of SIC-SCAN(complex) matches the CCSD(T) curve in both equilibrium (around $r \approx 0.8$ \AA) and stretched-bond ($r > 0.8$ \AA) regions. For comparison, the energy curve of SIC-SCAN(real) deviates from the one of CCSD(T) in the equilibrium region.

\begin{figure}[!htbp] 
    \begin{center}
            \includegraphics[width=3.4in, angle=0]{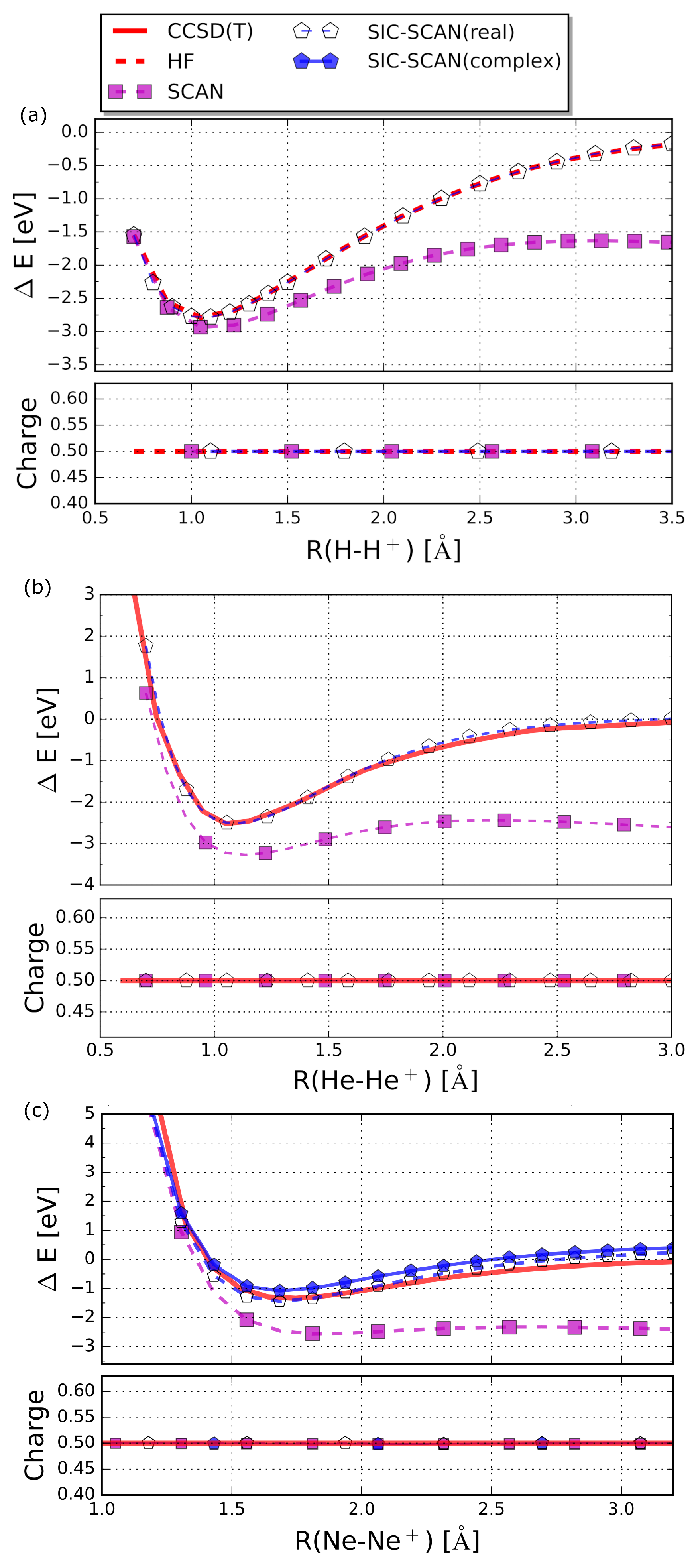}
  \end{center}
    \caption{ 
  Dissociation curves of H$_2^+$ (a), He$_2^+$ (b), and Ne$_2^+$ (c), with the zero-energy reference set to the total energy of isolated atoms/ions.
  The mirror symmetry ensures no charge transfer in these two dissociations, as demonstrated by the Mulliken charge analyses.
  All calculations were performed in \ti{FHI-aims}. \ti{Tight} \ti{tier}-1 NAO basis sets were employed for SCAN and SIC-SCAN incorporating \ti{real} and \ti{complex} SIC orbitals, while the CCSD(T) results utilize cc-pVTZ Gaussian-type basis sets. 
  The Hartree-Fock (HF) method, being exact for one-electron systems, provided the H$_2^+$  dissociation reference curve.
  }
  \label{Fig-dissociation-AA}
  \end{figure}

Fig.~\ref{Fig-dissociation-AA} illustrates how the SIE in SCAN calculations affects open-shell configurations, e.g., H$^+_2$, He$^+_2$ and Ne$^+_2$. For H$^+_2$, which contains only one electron, SIC exactly offsets the SIE in the SCAN calculation. However, for the He$^+_2$ and Ne$^+_2$ dissociations with more electrons, this exact cancellation does not occur. As the dissociation behaviors of He$^+_2$ and Ne$^+_2$ are similar, we take the Ne$^+_2$ system with more electrons as an example to analyze in the following. The conclusions drawn for Ne$^+_2$ are applicable to He$^+_2$ as well.
Nevertheless, SIC, whether using \ti{real} or \ti{complex} SIC orbitals, significantly improves the SCAN curve. The resulting SIC-SCAN curves align closely with CCSD(T) calculations, indicating that the PZ-SIC scheme successfully rectifies the majority of the many-electron SIEs in the SCAN calculations of the Ne$^+_2$ dissociation. 
As shown in Fig.~\ref{Fig-dissociation-AA}, both SCAN and SIC-SCAN, without symmetry breaking, predict the same charge distribution, with each Ne atom carrying a fractional charge of 0.5 (Ne$^{0.5}$).
It is also worth noting the combined total energy of Ne and Ne$^{+1}$, which is taken as the reference for the Ne$^+_2$ dissociation curve in Fig.~\ref{Fig-dissociation-AA}.
Regardless of the chosen SIC orbitals (either \ti{real} or \ti{complex}), SIC-SCAN closely reproduces the degeneracy between the configurations of Ne$^+ \cdots $Ne and Ne$^{0.5} \cdots $Ne$^{0.5}$, with a deviation of less than 0.3 eV. This observation suggests that SIC-SCAN can provide an accurate description of fractional charged systems, which has been widely used to discuss the SIE in the conventional DFAs \cite{cite-SIC-27-Yang-LOSC}.

\subsection{Periodic Systems}

\subsubsection{Performance for Crystal Properties}
Given that our SIC-SCAN implementation effectively addresses the multiple solution issue in both finite and period systems,
we have extended our attention to the crystal properties, including cohesive energy, bulk modulus, and lattice constant. 
\begin{figure}[!htbp] 
    \begin{center}
  \includegraphics[width=3.4in,angle=0]{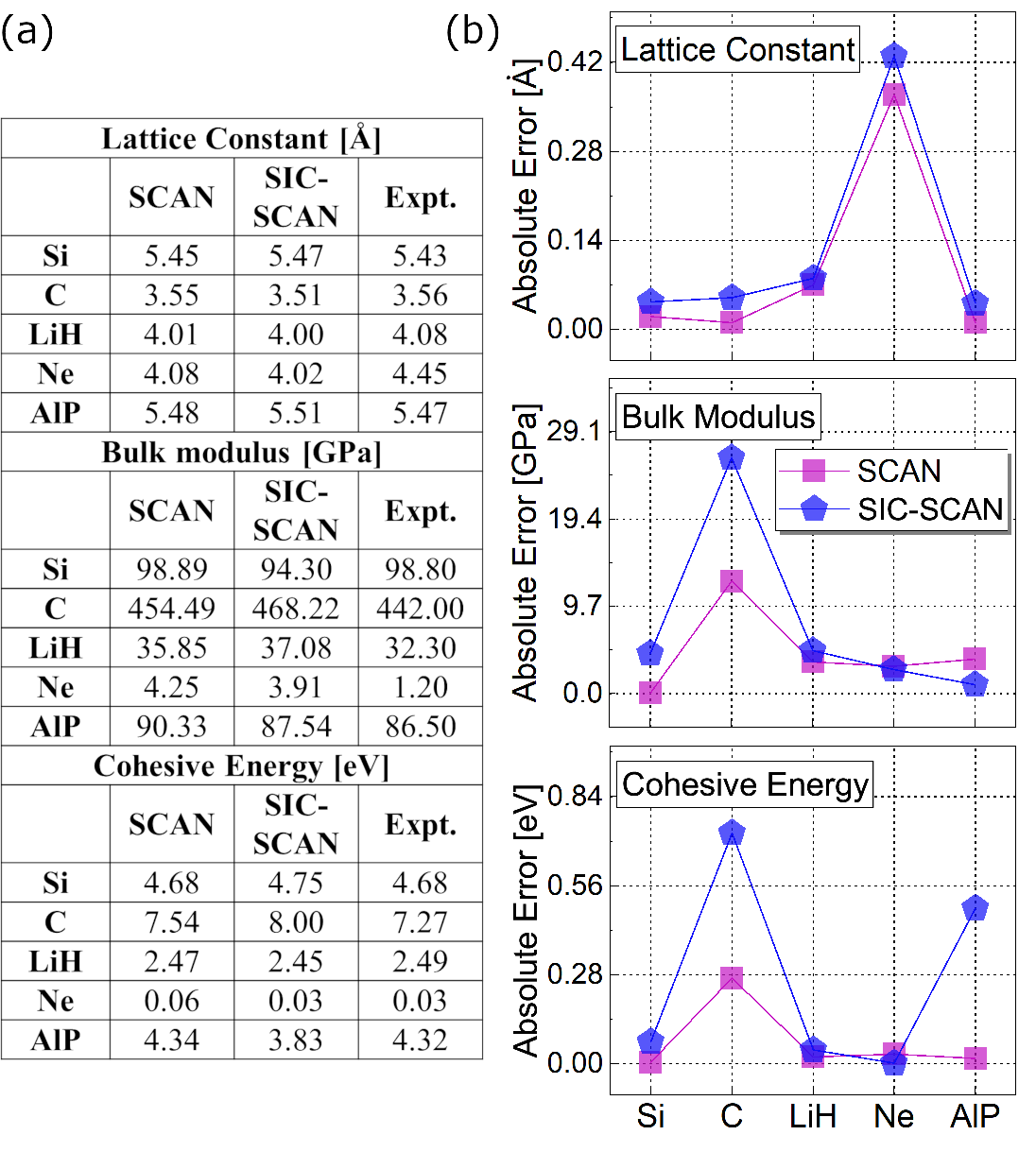}
  \end{center}
    \caption{ 
    (a) Lattice constants at equilibrium (\AA), bulk modulus (GPa) and cohesive energies (eV) of the diamond Si, diamond C, rocksalt LiH, fcc Ne, and zincblende AlP, calculated using SCAN and SIC-SCAN calculations. 
    (b) Absolute errors in the calculated lattice constants (\AA), bulk modulus (GPa), and cohesive energies (eV) for SCAN and SIC-SCAN. 
    All calculations were carried out in \ti{FHI-aims} using \ti{tight} \ti{tier}-1 NAO basis sets and the $\tb{k}$-grid setting of $4\times4\times4$. 
    The experimental reference values were collected from the main-group test set \cite{cite-MSE,cite-ddhybrid-MSE}. 
            }
  \label{Fig-eos}
  \end{figure}
We evaluate the cohesive energies of a fundamental set of materials, including diamond Si, diamond C, rocksalt LiH, fcc Ne, and zincblende AlP, as presented in Fig.~\ref{Fig-eos}-(a). Additionally, we computed the lattice constant and bulk modulus for this test set, with results also displayed in Fig.~\ref{Fig-eos}-(a). The relevant absolute errors (AEs) are plotted in Fig.~\ref{Fig-eos}-(b) for comparative analysis. 

Consistent with findings in previous studies \cite{cite-meta-problem-test,cite-SIC-SCAN}, our calculations suggest that the SIC contribution tends to deteriorate the accuracy of SCAN predictions for cohesive energies and bulk modulus. However, the impact of SIC on SCAN's lattice constant predictions is relatively minor. As a typical system of van der Waals (vdW) interaction, the fcc Ne poses a challenge for standard DFAs. Our results suggest that the SCAN method tends to overestimate the vdW interaction, with its calculated cohesive energy being double the experimental value and its lattice constant being 0.37 \AA~shorter. Given that the PZ-SIC scheme is irrelevant to addressing the SCAN error for vdW interactions, SIC-SCAN does not improve the terrible performance of SCAN in predicting the lattice constant of fcc Ne. The observed improvements in the cohesive energy and bulk modulus of the fcc Ne might be due to a fortuitous offset between the inappropriate SIC contribution and SCAN's overestimation of vdW interactions.

\subsubsection{Band Gaps}
\begin{figure}[!htb] 
    \begin{center}
  \includegraphics[width=3.2in,angle=0]{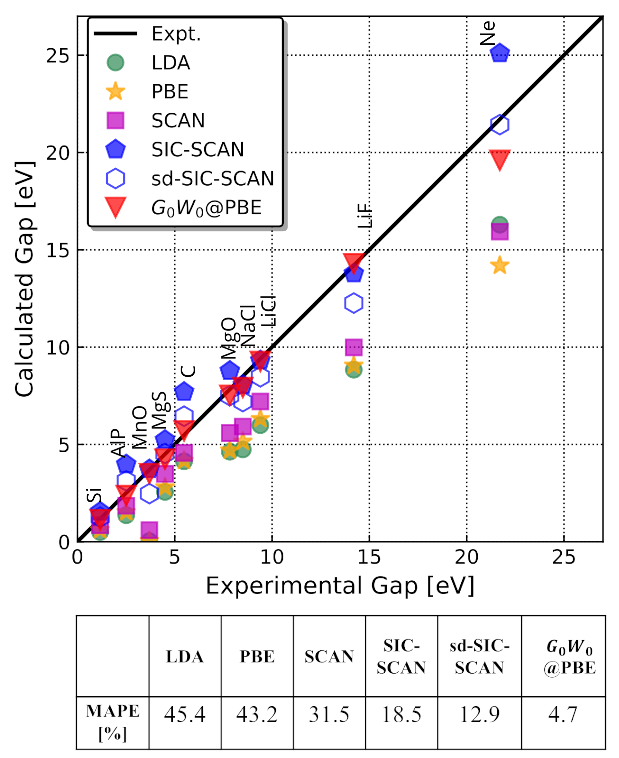}
  \end{center}
    \caption{ 
    Calculated band gaps using LDA, PBE, SCAN, (sd-)SIC-SCAN, and $GW$ against experimental counterparters. The graph features a black diagonal line representing the perfect agreement to the experiment. Mean absolute percentage errors (MAPEs) are collected below the graph. 
    Calculations of LDA, PBE, SCAN, and (sd-)SIC-SCAN were performed in \ti{FHI-aims}. 
    A $\tb{k}$-grid setting of $5\times5\times5$ was used together with \ti{tight} \ti{tier}-1 NAO basis sets. 
        Geometries were taken from the experiments. 
    Results for $G_0W_0$@PBE and experimental data were collected from Refs.~\cite{cite-GW-gap, cite-GW-LiCl}. 
      }
  \label{Fig-scaled-SIC-gap}
  \end{figure}

The fundamental band gap is a ground-state property, which is defined as the difference between the ionization energy and the electron affinity \cite{cite-SIC-26-Yang-GSC}. For periodic solids, it has been demonstrated that, for any DFA within the KS or gKS framework, the fundamental band gap is equivalent to the energy difference between the lowest unoccupied orbital (LUMO) and HOMO, providing that the given DFA potential is continuous upon introducing an electron or hole \cite{cite-SIC-26-Yang-GSC,cite-scan-gKS-solid}. This finding is consistent across both the KS and gKS methods. The well-documented underestimation of the band gap in semi-local DFAs, including LDA, PBE, and SCAN, is largely attributed to significant SIEs present in these approximations \cite{cite-GEA-2, cite-SIC-26-Yang-GSC}.
In this work, we evaluate the performance of various methods in predicting band gaps for a selection of 10 fundamental solids. This includes covalent crystals (such as Si, C), ionic crystals (such as NaCl, LiF), and Mott insulators composed of transition-metal oxides (MnO). The experimental band gaps are taken as the reference. 
It's worth emphasizing that although electron-phonon interactions can influence experimental band gaps \cite{many-particle-pyhsics,e-ph-anharmonic-band, e-ph-gap-C10H8}, 
the remarkable agreement between the band gaps determined by the $G_0W_0$ method and experimental data suggests that, at least for the solids investigated here, the impact of this interaction is probably less significant compared to the SIE in DFAs.
As illustrated in Fig.~\ref{Fig-scaled-SIC-gap}, the mean absolute percentage errors (MAPEs) of LDA and PBE are 45.4~\% and 43.2~\%, respectively, substantially higher than the $G_0W_0$ value (4.7~\%). SCAN outperforms LDA and PBE with a lower MAPE (31.5~\%), as the step outside KS to gKS is to enhance orbital energies. Unfortunately, However, Figure \ref{Fig-scaled-SIC-gap} reveals that SCAN still has poor accuracy in predicting band gaps, consistent with previous findings~\cite{cite-scan-3-solid}. Our results suggest that applying SIC to SCAN can reduce the band gap error, lowering the MAPE to 18.5~\%. We acknowledge that SIE may not be the sole source of error for SCAN, but it significantly contributes to inaccuracies in band gap calculations.

MnO is classified as a Mott insulator characterized by strong electron interactions~\cite{cite-Mott, cite-Hubbard}. Accurately predicting the band gap of MnO is a challenge for conventional semi-local DFAs~\cite{cite-SIE-Mott}. Our results confirm that LDA, PBE, and SCAN methods predict the band gaps for MnO around 0.60 eV, a significant underestimation compared to the experimental value of 3.70 eV. 
A common remedy is the DFT$+U$ method, which introduces an empirical electron repulsion ``$U$'' \cite{cite-LDA-U, cite-GW-LDA-U}.
For instance, GGA$+U$ gives a band gap of approximately 4 eV when applying a ``$U$'' of 5.25 eV via the linear-response approach \cite{cite-GGA-U-linear}. Moreover, the self-consistent $GW_0$ approach, based on the LDA$+U$ wave functions, produces a band gap of 3.32 eV \cite{cite-revGW-LDA-U}. In contrast, our research indicates that the PZ-SIC scheme can also effectively improve the SCAN description for MnO's band gap. The SIC-SCAN method calculates a band gap of 3.73 eV, which is impressively close to the experimental value, deviating by a mere 0.03 eV. 
This result aligns with previous studies, suggesting that PZ-SIC captures the strong electron interactions often overlooked in standard semi-local DFAs~\cite{cite-SIC-4-manybody, cite-SIE-Mott}. 

To further improve the calculated band gaps in the KS-DFT framework, a scaled-down PZ-SIC (sd-SIC) variant has been proposed~\cite{cite-SIC-10-scaldown, cite-SIC-fermi-solid, cite-FSIC-fractional-scaldown}. Compared to the standard PZ-SIC scheme (Eq.~\ref{eq:SIC}), the pivotal adjustment in sd-SIC is to introduce a set of orbital-density specific scaling factors, denoted as $\{X_{i}\}$, into the self-interaction correction 
\begin{equation} \label{eq:sd-energy}
  \begin{aligned}
      & E^{\tf{sd-SIC-DFA}} = E^{\tf{DFA}} + E^{\tf{sd-SIC}} \\
      & E^{\tf{sd-SIC}}[\{n_i\}] = - \sum_i^{N_{\tf{e}}} X_{i} \left( E_{\tf{xc}}^{\tf{DFA}}[n_{i}]+E_{\tf{H}}[n_{i}] \right) \quad.
  \end{aligned}
\end{equation}
In addition to determining the scaling factor for each orbital, adopting a uniform scaling factor across all orbital densities ($X_{i} = X$) presents a straightforward and effective strategy, albeit it takes on empirical impacts.
For instance, it has been observed that a scaling factor of $0.4$ enhances thermochemistry \cite{cite-SIC-10-scaldown}, while a factor of $0.2$ yields desirable band gaps in both silicon and diamond \cite{sd-SIC-02}. 
Furthermore, the coeﬀicient $X=2/3$, as proposed by Shinde \etalcite~\cite{cite-SIC-fermi-solid} in 2020, to enhanced the SIC-PBE results for fundamental solids. Building upon this, our study investigates the impact of this coefficient on the SIC-SCAN band gaps. We discovered that a scaling factor of 0.60 yields optimal results for SIC-SCAN. These findings are comprehensively discussed in Supplementary Section V. The outcomes are illustrated in Fig.~\ref{Fig-scaled-SIC-gap}, where they are denoted as scaled-down SIC-SCAN (sd-SIC-SCAN).
In agreement with the previous research by Shinde \etalcite~\cite{cite-SIC-fermi-solid}, 
we find that the scaled-down algorithm with a constant scaling factor enhances the accuracy of the calculated band gaps.
The sd-SIC-SCAN approach outperforms the SIC-SCAN method, bringing down the MAPE to 12.9\%. 
However, it is crucial to highlight that the standard PZ-SIC method is exact for one-electron systems, 
which, unfortunately, is not the case in the sd-SIC method.

\subsubsection{Crystalline \ti{trans}-Polyacetylene}

\begin{figure*}[!htbp] 
      \begin{center}
    \includegraphics[width=5.8in,angle=0]{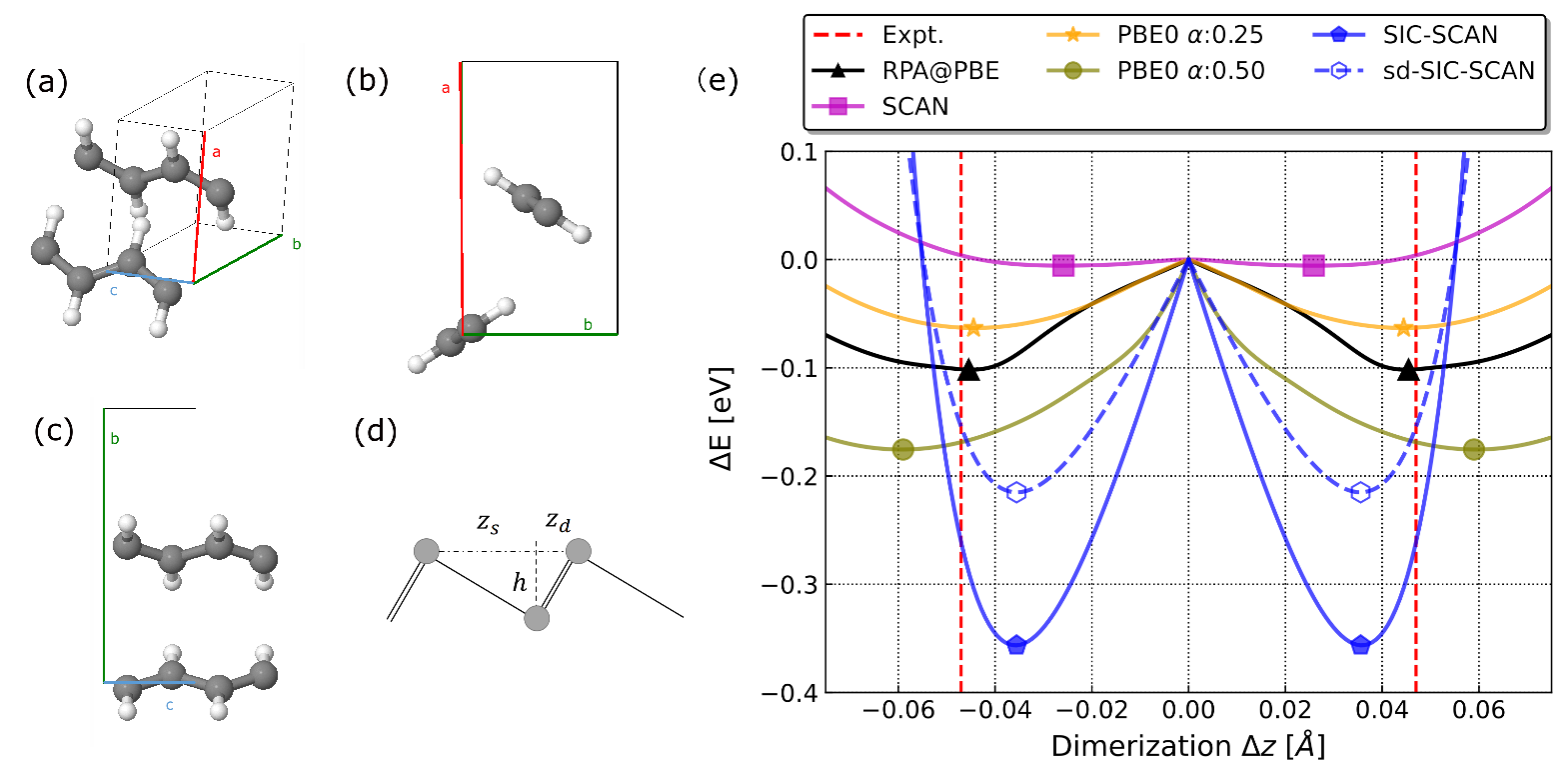}
    \end{center}
        \caption{ 
      (a)  Experimental unit cell of the TPA structure~\cite{cite:TPA-xray-2}. Carbon atoms are depicted in gray, while hydrogen atoms are in white. 
      (b) and (c) are the right and bottom side views, respectively.
      (d) The schematic view of a single (CH)$_x$ chain with the dimerization parameter $\Delta z$, representing the difference between the distance of single-bond and double-bond carbon atoms along the chain direction $z_s$ and $z_d$. 
      (e) Total energy curves as a function of the dimerization parameter $\Delta z$, computed using the SCAN, (sd-)SIC-SCAN, PBE0, and RPA@PBE methods. 
      All calculations were executed in \ti{FHI-aims}. 
      Vertical red dashed lines in the subfigure (e) mark the experimental dimerization parameter ($\Delta z \approx 0.047$ \AA{}) \cite{cite:TPA-xray}.
      The calculated total energies for $\Delta z =0$ were set as the zero-energy reference.
      Markers pinpointed the optimized dimerization for each method.
            A dense $\tb{k}$-grid setting of $14\times12\times16$ was used together with \ti{tight} \ti{tier}-1 NAO basis sets for these calculations. 
            As the dimerization factor varies, the combined length of $z_s + z_d$ remained equal to the lattice vector length $|c|$. A vertical height $h$ of $0.70$ \AA{}--from the middle carbon atom to the connection of the other two carbon atoms--was maintained.
    }
    \label{Fig-dimer}
  \end{figure*}
We then turn to a challenging problem for periodic systems, the broken symmetry in crystalline polyacetylene (CH)$_x$. The polyacetylene chain has two configurations, named \textit{cis} and \textit{trans}. In 1957, Ooshika found that the \ti{trans} configuration is a stable configuration for the long-chain polyacetylene (TPA) \cite{cite-SSH-expt, cite-polyacetylene-1, cite-polyacetylene-2},
where a sequence of alternating single and double carbon bonds is formed along the chain, as illustrated in Fig.~\ref{Fig-dimer}-(a-d). 
The dimerization parameter $\Delta z$ describes the length difference of these bonds along the chain's direction ($z_s$ and $z_d$):
\begin{equation}
  \begin{aligned}
    \Delta z = z_s - z_d \quad.
  \end{aligned}
\end{equation}
Here, $z_s$ and $z_d$ are vertical distances from the central carbon atom on both sides. X-ray studies on crystalline TPA identified a dimerization of $\Delta z \approx 0.047$ \AA{}~\cite{cite:TPA-xray}. 
The underlying mechanism of symmetry break is yet to be clarified. On the one hand, the formation of alternating bonds and charge transfer between carbons has been explained by using the celebrated 1D Su-Schrieffer-Heeger model, emphasizing the phonon/electron-phonon interaction~\cite{cite-SSH}.
On the other hand, the electron-electron interaction could be another possible driving force for the induced dimerization~\cite{cite:TPA-b3lyp,cite-SIC-4-hybrid, cite:pbe0}.

Fig.~\ref{Fig-dimer}-(e) displays the TPA energy curves as a function of the dimerization factor $\Delta z$, which are calculated using the SCAN, SIC-SCAN, PBE0, and RPA@PBE methods. In this context, ``RPA@PBE'' denotes the random phase approximation based on PBE density and orbitals\cite{cite-RPA-aims}.
The standard hybrid functional PBE0 \cite{cite:pbe0}, which incorporates a portion of the Hartree-Fock exchange with $\alpha=0.25$, predicts a dimerization of 0.044 \AA{}. This aligns closely with the experimental findings (0.047 \AA{}) and the RPA@PBE calculation (0.045\AA{})~\cite{cite-RPA-aims}. Our analysis indicates that the calculated dimerization is sensitive to the $\alpha$ value. When $\alpha$ is increased to 0.50, the resulting dimerization parameter rises to 0.059 \AA{}, overshooting the experimental value by 0.012 \AA.

While all the methods investigated here predict a symmetry-broken \textit{trans} configuration with $\Delta z>0$, the dimerization factor $\Delta z$ determined by the SCAN method is 0.026 \AA{}, significantly shorter than the experimental measurement of 0.047 \AA{}. 
In agreement with a previous study~\cite{cite:TPA-SIE}, our results confirm that the PZ-SIC scheme enhances the accuracy of conventional semi-local DFAs for this system. SIC-SCAN predicts a more extended dimerization parameter of around 0.036 \AA{}. While the calculated dimerization parameter remains unchanged when applying a uniform scaling factor $X=0.60$ via the scaled-down SIC formula (Eq.~(\ref{eq:sd-energy})), the sd-SIC-SCAN method considerably diminishes the energy well compared to the SIC-SCAN method.

                                                                    \begin{figure}[!htbp] 
            \begin{center}
      \includegraphics[width=3.2in,angle=0]{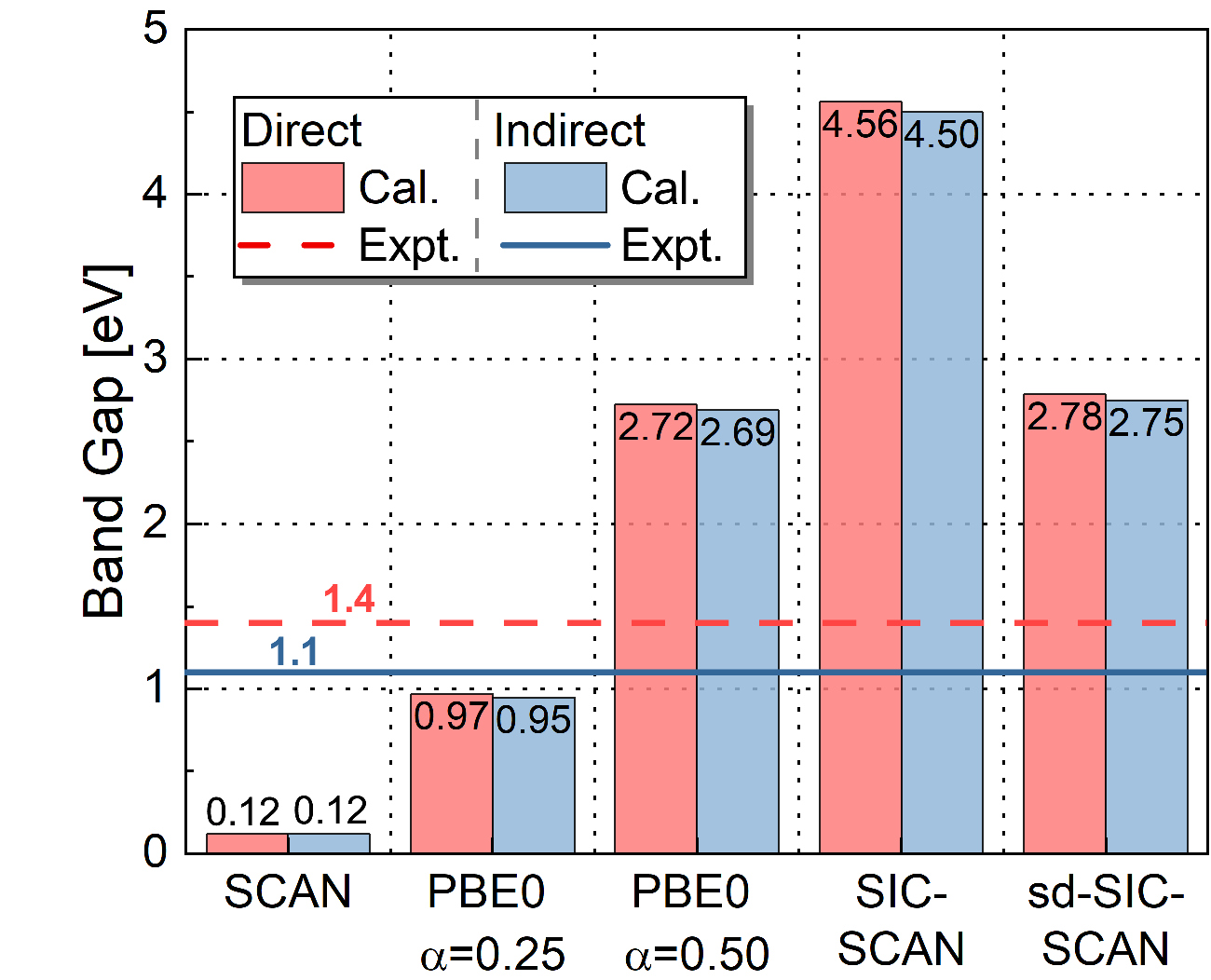}
      \end{center}
            \caption{Calculated band gaps of crystalline \ti{trans}-polyacetylene using the SCAN, PBE0 and \text{SIC-SCAN} methods.
      All calculations were performed in \ti{FHI-aims}. The experimental geometry was taken from Ref.~\cite{cite:TPA-xray}.
            A dense $\tb{k}$-grid setting of $14\times12\times16$ was used together with the \ti{tight} \ti{tier}-1 NAO basis sets. 
      The blue dash line and orange line presented the direct and indirect experimental band gaps, respectively,
      which were taken from Ref.~\cite{cite:TPA-xray}.
      }
      \label{Fig-CH-gap}
          \end{figure}

Finally, we turn our attention to the band gap of crystalline TPA, using its experimental geometry from reference~\cite{cite:TPA-xray}. Direct and indirect adsorption gaps of TPA have been identified by Fincher\etalcite, which are 1.4 eV and 1.1 eV, respectively~\cite{cite:TPA-xray}. 
Fig.~\ref{Fig-CH-gap} presents the direct and indirect band gaps as calculated by various methods. 
The standard hybrid PBE0 with $\alpha=0.25$ slightly underestimates the indirect band gap at 0.95 eV. In contrast, using $\alpha=0.50$ with PBE0 leads to a notable overestimation.
As expected, due to the serious SIE, SCAN significantly underestimates the band gap, erroneously predicting identical values of 0.12 eV for both direct and indirect gaps. The SIC-SCAN method, on the other hand, rectifies the difference between direct and indirect gaps but greatly overestimates the indirect band gap to 4.50 eV. Introducing a uniform scaling factor, $X$, alleviates the overcorrection of PZ-SIC, but the calculated indirect band gap of sd-SIC-SCAN remains considerably larger than the experimental findings. 

Crystalline TPA represents a challenging periodic system to study. Numerous factors, including lattice constants~\cite{cite-polyacetylene-stress}, can influence its calculated properties. 
As illustrated in Fig.~\ref{Fig-dimer} and Fig.~\ref{Fig-CH-gap}, significant discrepancies are evident in energy and band gap values across various computational methods. The widely used hybrid functional PBE0, with a default alpha of $0.25$, predicts dimerization that closely aligns with the results of RPA@PBE. However, increasing the alpha value to 0.5 adversely affects PBE0's performance. It's worth noting that PBE0 outperforms the (SIC-)SCAN method in predicting these TPA properties, highlighting that while mitigating the SIE is crucial, it is not the only factor to consider when analyzing crystalline TPA using density functional theory.

\section{Conclusion} \label{sec-conclusion}

The state-of-the-art, non-empirical meta-GGA SCAN functional provides a significant and consistent improvement over the non-empirical GGA PBE across a range of chemical interactions in both molecules and solids. However, it is not immune to the pervasive self-interaction error.
In this study, we implement and generalize the self-interaction correction algorithm proposed by Perdew and Zunger (PZ-SIC), to alleviate this error within the SCAN functional. Specifically, we introduce the orbital density-potential constraint, which, in conjunction with the total density constraint and the orbital potential constraint, facilitates the self-consistent localization of the SIC orbitals, leading to an accurate PZ-SIC energy.
Our findings indicate that the orbital density-potential constraint can narrow down multiple potential SIC orbitals to a singular optimal choice in systems like C, He$_2$, hcp-Helium, and the CH-chain.

Our PZ-SIC approach markedly enhances the performance of SCAN, as evidenced by the highest occupied molecular orbital (HOMO) energies $\epsilon_{\tf{HOMO}}$ of H-Ar atoms and a set of 18 molecules. Built upon the accurate description of orbital energies, our PZ-SIC approach yields commendable energy curves and corrected charge transfers in the dissociations of H-F and Ne-Ne$^+$.  
When tested on 10 foundational solids, SIC-SCAN consistently outperforms SCAN in band gap predictions. However, for other crystalline properties, such as cohesive energy, bulk modulus, and lattice constants, SIC-SCAN does not fare as well as SCAN for our test set, which includes diamond Si, diamond C, rocksalt LiH, fcc Ne, and zincblende AlP. We then delve into the ground state of crystalline polyacetylene, a renowned challenge for periodic systems that remains elusive for semi-local DFT calculations like the SCAN functional. Our findings indicate that the PZ-SIC methodology adeptly identifies the symmetry-breaking in crystalline polyacetylene, emphasizing the role of electron-electron interactions.

\section*{supplementary Information}
The following files are available free of charge: 
\begin{itemize}
	\item supplementary.pdf, containing additional information needed for detailed discussions.
	\item data.zip, containing geometries and electronic-structure energy results produced in this project and also available on the NOMAD repository with the DOI: 10.17172/NOMAD/2023.11.28-1 ~.
\end{itemize}
\begin{acknowledgments}
	Sheng thanks Zhenkun Yuan, Hagen-Henrik Kowalski and Florian Knoop for inspiring discussions. 
	This work received funding from the European Union's Horizon 2020 Research and Innovation Programme (grant agreement No.~951786, the NOMAD CoE), and the ERC Advanced Grant TEC1P (No.~740233).
	This work was partially supported by the National Natural Science Foundation of China (No.~21973015, No.~22125301) and Innovative Research Team of High-Level Local universities in Shanghai, and a Key Laboratory Program of the Education Commission of Shanghai Municipality (ZDSYS14005).
\end{acknowledgments}
\bibliographystyle{apsrev4-1}
\bibliography{apssamp}
\end{document}